%% file: main.tex
\RequirePackage{fix-cm}
\documentclass[smallextended]{svjour3}

\input{setup}

\begin{document}

\title{An Empirical Study of Token-based Micro Commits}
\titlerunning{An Empirical Study of Token-based Micro Commits}

\author{Masanari Kondo \and
        Daniel M. German \and
        Yasutaka Kamei \and
        Naoyasu Ubayashi \and
        Osamu Mizuno
}

\institute{
    Masanari Kondo, Naoyasu Ubayashi\at
    Principles of Software Languages group (POSL)\\
    Kyushu University, Japan\\
    \email{\{kondo, ubayashi\}@ait.kyushu-u.ac.jp}\\\\
    Daniel M. German\at
    Department of Computer Science, University of Victoria,\\
    Victoria, BC, Canada\\
    \email{dmg@uvic.ca}\\\\
    Yasutaka Kamei\at
    Principles of Software Languages group (POSL)\\
    Kyushu University, Japan, and\\
    Inamori Research Institute for Science, Kyoto, Japan\\
    \email{kamei@ait.kyushu-u.ac.jp}\\\\
    Osamu Mizuno\at
    Software Engineering Laboratory (SEL)\\
    Kyoto Institute of Technology, Japan\\
    \email{o-mizuno@kit.ac.jp}\\\\
}

\authorrunning{Kondo~\et{}} 

\maketitle

\makeatletter
\def\ps@headings{%
\def\@oddfoot{\scriptsize \hfill \thepage }%
\def\@evenfoot{\scriptsize \thepage \hfill}}
\makeatother
\pagestyle{headings}

\input{sections/abstract}
\input{sections/introduction}
\input{sections/motivation}

\input{sections/dataset}

\input{sections/rq1}
\input{sections/rq2}
\input{sections/rq3}
\input{sections/discussion}
\input{sections/threats}
\input{sections/related}
\input{sections/conclusion}

\section*{Acknowledgment}
We gratefully acknowledge the financial support of: (1) JSPS for the KAKENHI grants (JP21H04877, JP22K18630, JP22K17874, JP24K02921), Bilateral Program grant JPJSBP120239929; (2) the Inamori Research Institute for Science for supporting Yasutaka Kamei via the InaRIS Fellowship; and (3) NSERC Canada.

\section*{Declarations}
\subsection*{Funding and/or Conflicts of interests/Competing interests}
All authors certify that they have no affiliations with or involvement in any organization or entity with any financial
interest or non-financial interest in the subject matter or materials discussed in this manuscript.

\subsection*{Data Availability Statements}
The replication package that supports the findings of this study is available.\footnote{\url{https://doi.org/10.5281/zenodo.10963270}}


\bibliographystyle{spmpsci}
\bibliography{main}

\include{sections/appendix}

\end{document}

%% file: setup.tex
\usepackage{adjustbox}
\usepackage{afterpage}
\usepackage{balance}
\usepackage{booktabs}
\usepackage{caption}
\usepackage{cite}
\usepackage{colortbl}
\usepackage{collectbox}
\usepackage{comment}
\usepackage{enumitem}
\usepackage{framed}
\usepackage{graphicx}
\usepackage{lscape}
\usepackage{lipsum}
\usepackage{listings}
\usepackage{longtable}
\usepackage{multirow}
\usepackage{tabularx}
\usepackage{txfonts}
\usepackage{url}
\usepackage{varwidth}
\usepackage{xspace}
\usepackage[dvipsnames]{xcolor}
\usepackage[tight,footnotesize]{subfigure}
\usepackage{tcolorbox}
\tcbuselibrary{raster,skins}
\definecolor{light-gray}{gray}{0.85}

\newcommand{\countobservations}{
    \def \countobservations{1}
}
\newcounter{observation}
\countobservations

\newcommand{\countmess}{
    \def \countmess{1}
}
\newcounter{mes}
\countmess

\def\summarybox#1#2{
\medskip
\begin{tcolorbox}[enhanced,title=#1, attach boxed title to top left= {xshift=0mm, yshift*=-\tcboxedtitleheight/2}]
    #2
\end{tcolorbox}
}

\newlist{stepitemize}{itemize}{1}
\setlist[stepitemize,1]{leftmargin=1.26cm}

\definecolor{diffstart}{named}{Gray}
\definecolor{diffincl}{named}{Green}
\definecolor{diffrem}{named}{Purple}
\lstdefinelanguage{diff}{
basicstyle=\ttfamily\small,
float,
floatplacement=t,
frame=tb,
morecomment=[f][\color{diffstart}]{@@},
morecomment=[f][\color{diffincl}]{+},
morecomment=[f][\color{diffrem}]{-},
}

\lstdefinelanguage{difftab}{
basicstyle=\ttfamily\scriptsize,
breaklines=true,
morecomment=[f][\color{diffstart}]{@@},
morecomment=[f][\color{diffincl}]{+},
morecomment=[f][\color{diffrem}]{-},
}

\newcommand{\reb}[1]{\textcolor{red}{#1}}

\def\et{et\ al.\xspace}
\def\ie{i.\,e.\xspace}
\def\eg{e.\,g.\xspace}

\def\fig#1{Figure~\ref{#1}}

\def\tab#1{Table~\ref{#1}}

\def\sec#1{Section~\ref{#1}}
\def\lis#1{Listing~\ref{#1}}
\def\liss#1{Listings~\ref{#1}}

\newlength\MAX  
\newcommand*\Chart[2]{#1~\rlap{\rlap{\textcolor{black!20}{\rule{\MAX}{2ex}}}\rule{#2\MAX}{2ex}}\hspace{6.5mm}{\ }}

\newcommand{\cregit}{\texttt{cregit}\xspace}

\newcommand{\oneLine}{one-line commit\xspace}
\newcommand{\oneLines}{one-line commits\xspace}

\newcommand{\sqlite}{SQLite\xspace}

\usepackage{framed}
\usepackage{xparse}
\newsavebox{\fminipagebox}
\NewDocumentEnvironment{hassanbox}{m O{\fboxsep}}
 {\par\kern#2\noindent\begin{lrbox}{\fminipagebox}
  \begin{minipage}{#1}\ignorespaces}
 {\end{minipage}\end{lrbox}%
  \makebox[#1]{%
    \kern\dimexpr-\fboxsep-\fboxrule\relax
    \fbox{\usebox{\fminipagebox}}%
    \kern\dimexpr-\fboxsep-\fboxrule\relax
  }\par\kern#2
 }

\newcommand{\microCs}{micro commits\xspace}

\newcommand{\RQone}{What are the characteristics of \microCs?}
\newcommand{\RQtwo}{What are the types of changes that \microCs perform?} 
\newcommand{\RQthree}{How do micro commits compare to \oneLines{}?}

\newcommand{\RQCone}{What Are the Characteristics of Micro Commits?}
\newcommand{\RQCtwo}{What Are the Types of Changes That Micro Commits Perform?} 
\newcommand{\RQCthree}{How Do Micro Commits Compare to One-line Commits?}


%% file: sections/abstract.tex
\begin{abstract}

In software development, developers frequently apply maintenance activities to the source code that change a few lines by a single commit.
A good understanding of the characteristics of such small changes can support quality assurance approaches (e.g., automated program repair), as it is likely that small changes are addressing deficiencies in other changes; thus, understanding the reasons for creating small changes can help understand the types of errors introduced. Eventually, these reasons and the types of errors can be used to enhance quality assurance approaches for improving code quality.
While prior studies used code churns to characterize and investigate the small changes, such a definition has a critical limitation. 
Specifically, it loses the information of changed tokens in a line. 
For example, this definition fails to distinguish the following two one-line changes: (1) changing a string literal to fix a displayed message and (2) changing a function call and adding a new parameter.
These are definitely maintenance activities, but we deduce that researchers and practitioners are interested in supporting the latter change. 
To address this limitation, in this paper, we define \emph{micro commits}, a type of small change based on changed tokens.
Our goal is to quantify small changes using changed tokens.
Changed tokens allow us to identify small changes more precisely. 
In fact, this token-level definition can distinguish the above example.
We investigate defined micro commits in four OSS projects and understand their characteristics as the first empirical study on token-based micro commits.
We find that micro commits mainly replace a single name or literal token, and micro commits are more likely used to fix bugs. 
Additionally, we propose the use of token-based information to support software engineering approaches in which very small changes significantly affect their effectiveness. 

\keywords{Empirical Study; Micro Commits; Tokens; Mining Software Repositories}
\end{abstract}

%% file: sections/introduction.tex
\section{Introduction}
\label{sec:introduction}

Commits changing a few lines of code are common in software development. Purushothaman et al. defined \emph{small commits} as those modifying less than 10 lines in their study~\cite{purushothaman2005TSE}. They found that 50\% of changes in the examined systems were small commits.
They also reported that 10\% of all commits were \emph{\oneLines{}} (modified at most one line). 
In a recent study, Alali et al. reported that in the GCC project, 19.9\% of commits were \emph{extra-small}, adding at most 5 lines of code~\cite{alali2008ICPC}.
Our research found that in the projects we studied, between 6 and 8\% of all commits were \oneLine{}s (see \sec{sec:motivation}).


Improving code quality is an ultimate goal for software engineering researchers, and several quality assurance
approaches have been widely studied so far, such as automated program repair (APR), defect prediction, and fault
localization.  A good understanding of the characteristics of very small changes can support such approaches, as it is likely
that such changes are addressing deficiencies in the system~\cite{purushothaman2005TSE}; thus, understanding
the characteristics of creating small changes can help understand the types of errors that other changes introduce
and potentially help with program repair.
Eventually, the information can be used to enhance quality assurance approaches for improving code
quality.

While prior studies~\cite{purushothaman2005TSE,hindle2008MSR,alali2008ICPC} use churn (number of lines added and
removed) to identify small changes (\eg, small commits and \oneLines{}~\cite{purushothaman2005TSE}, or extra-small commits~\cite{alali2008ICPC}), it has one significant limitation: they consider the line to be the finest-grained entity of changed source code.
More specifically, such a definition overlooks the details of what has changed in a line~\cite{german2019EMSE,meng2013ICSM,servant2017ICSE}.
For instance, when several lines have a small change (such as an identifier being renamed in a few places), these modifications might appear as one line added and one line removed for each change, rather than a single identifier change.


Another problem is that splitting or joining a line of code that is being modified can result in noise.
For example, splitting a line into two would be reflected as a change to multiple lines in version control systems
(\eg, Git), and this type of change can add noise to the analysis of the history of the development process. 

These limitations sometimes cause researchers to fail in accurately quantifying small changes.
For example, \lis{code:ex-missed-micro-commit} shows a commit in the Linux repository that changes a few lines (\ie, three added and two deleted lines). 
While this commit corresponds with a multiple-line change and may not correspond to a \oneLine{}, it only adds a token ``static''.
This is similar to \lis{code:ex-micro-commit} corresponding with a \oneLine{} that only adds a token ``static''.
Studying the actual changed tokens instead of the lines can provide a better understanding of the characteristics of the small changes.


In this paper, we define a new class of commits: \emph{micro commits}.
\textbf{Micro commits are commits that add at most five tokens and remove at most five tokens of source code}.
We aim to quantify small changes using the token-level definition (\ie, micro commits) rather than relying on the line-level definition (\ie, \oneLines{}).
This token-level definition allows us to identify small changes more accurately, and use token information to
characterize them.
We conducted an empirical study on four large, mature open-source projects
to: a) demonstrate that micro commits are common, accounting for between 7.45 and 17.95\% of all studied commits in
the studied projects, b) understand their qualitative and quantitative characteristics, and c) show our definition of
micro commits (a threshold of 5 added and removed tokens) includes approximately 90\% of all \oneLines{}, yet only approximately 40--50\% of micro commits are \oneLine{}s.

Specifically, we answer the following research questions (RQs).
We have also provided a summary of the key findings for each RQ.
The detailed results are described in \sec{sec:rq1}, \sec{sec:rq2}, and \sec{sec:rq3}.

\begin{itemize}
\setlength{\leftskip}{0.3cm}
\item[RQ1:] \textbf{\RQone{}}
\\
\textit{Motivation:}
  This research question aims to explain their quantitative characteristics: how frequent they are, and the types of tokens they delete and add.
	\\
  \textit{Results:}
	Most micro commits replace a single token with one of the same types, and this token type is mostly name (\eg, identifier names) or literal (\eg, numbers).
	Java and C differ on the most frequent tokens in micro commits.
\item[RQ2:] \textbf{\RQtwo{}}
	\\
	\textit{Motivation:}
	We intend to understand the purpose of micro commits (\eg, changing control flow, replacing the name of a variable, and modifying an expression) and whether a micro commit performs one or more activities.
	Specifically, we manually inspected the changes applied to the source code to understand the purpose behind the micro commit and the occurrence of activities.
	\\
  \textit{Results:}
	More than 85\% of micro commits apply a single operation to a single target. 
	The four most common types of these micro commits are replacing an existing expression, identifier, constant, or declaration.
  Multi-operation micro commits usually change the order of statements.
\item[RQ3:] \textbf{\RQthree{}}
	\\
	\textit{Motivation:}
	This research question aims to explore the extent of differences between one-line commits and micro commits. Extracting micro commits requires syntactic parsing of the source code, which is more costly than extracting one-line commits. If they are identical, micro commits may be redundant.
	\\
  \textit{Results:}
	Most \oneLine{}s are micro commits (approximately 89--93\%).
	In contrast, only about 40--50\% of micro commits are one-line commits.
	Indeed, 30--40\% of micro commits include two or more hunks (one-line commits only have one hunk).
\end{itemize}

The main contributions of this paper are as follows:
\begin{itemize}
\item We propose the concept of micro commits as commits that add at most five tokens and remove at most five tokens, and
  demonstrate that these types of changes are common.  
\item We empirically investigate micro commits and understand their quantitative and qualitative characteristics.
We especially shed light on the differences in micro commits between programing languages through our manual inspection.
\item We propose the use of token-level information to support software engineering approaches that use extremely small changes (\eg, programing repair).
\item We provide the replication package of this study that contains a set of micro commits that have been manually labeled
  according to their purpose.
\end{itemize}

The organization of our paper is as follows:
\sec{sec:motivation} introduces motivating examples.
\sec{sec:dataset} explains our studied dataset. 
\sec{sec:rq1}, \sec{sec:rq2}, and \sec{sec:rq3} present the experiments and results based on our RQs.
\sec{sec:discussion} proposes the use of token-level information.
\sec{sec:threats} describes the threats to the validity of our case study.
\sec{sec:related} introduces related work.
\sec{sec:conclusion} presents the conclusion.

%% file: sections/motivation.tex
\section{Motivating Example}
\label{sec:motivation}
\input{tables/motivation/proportion-micro-commits}

In this section, we provide an example of a \oneLine{}. 
Also, we demonstrate that they account for a non-negligible proportion of commits.
Finally, we highlight the drawback of using lines of code to study extremely small changes, and we discuss how micro commits can address this drawback.

We first show the frequency of \oneLine{}s in four OSS projects used in this study and confirm that it is consistent with~\cite{purushothaman2005TSE}.
As in \cite{purushothaman2005TSE}, we use the diffs generated by Git to identify \oneLine{}s.
\tab{tab:motivation:micro-commits} shows the proportion of \oneLine{}s.
The proportion was computed by using the ``\#studied commits'' column. 
It only shows the commits that have made changes to the source code. Our analysis is conducted based on these commits. The detailed procedure for extracting commits is explained in \sec{sec:database:definition} and \sec{sec:dataset:collection}.\footnote{Because of the differences in source code management tools, one-line changes in the prior study~\cite{purushothaman2005TSE} and our \oneLine{}s are slightly different.}
We observe 4.28--8.20\% of \oneLine{}s. Specifically, the proportion is more than 7\% in the Linux and Zephyr projects.
Hence, \oneLine{}s account for a non-negligible proportion of all commits. 

\input{codes/motivation/ex-micro-commit}
\input{codes/motivation/ex-missed-micro-commit}

\lis{code:ex-micro-commit} shows an example of a \oneLine{}, also known as a micro commit.
This commit adds a static modifier into a struct definition, and this is not adding a functionality but fixing the code.

However, some extremely small changes are often obscured by splitting or joining lines of code, making them appear more complex than they are.
For example, \lis{code:ex-missed-micro-commit} shows an example of a micro commit that is not a \oneLine{}.
This commit semantically adds a static modifier only; however, this commit includes multiple changed lines because of changing the format of the definition of the variable.
\liss{code:ex-micro-commit} and~\ref{code:ex-missed-micro-commit} are semantically identical, but \oneLine{}s cannot include \lis{code:ex-missed-micro-commit} because it modifies multiple lines. 
Because we used Git, we deduced that the diff algorithms could address this limitation. 
Git has four algorithms to compute diffs, and they exhibit different results~\cite{nugroho2020dEMSE}.
Hence, we investigated four algorithms: \emph{patience}, \emph{minimal}, \emph{histogram}, and \emph{myers} described in the Git manual page.\footnote{\url{https://git-scm.com/docs/git-diff}}
However, all algorithms generate the same diff.
Hence, \oneLine{}s may overlook such commits.
If these diffs are analyzed with finer-grained source code entities (\eg, AST), it is easy to realize these commits have the same intention (\ie perform the same change).
However, AST analysis is expensive, particularly in repositories such as Linux that has more than one million commits and more than 60k source code files.

Therefore, to address this limitation, we define \textit{micro commits} based on tokens.
Because tokens are the semantically finest-grained source code entity, micro commits based on tokens can cover ones overlooked by \oneLine{}s.
Indeed, \liss{code:ex-micro-commit} and~\ref{code:ex-missed-micro-commit} change one token only; thus, they both perform
the same change in two different lines of code.

%% file: tables/motivation/proportion-micro-commits.tex

\begin{table}
    \caption{
        The proportion of \oneLine{}s in the studied projects 
}
\label{tab:motivation:micro-commits}
\begin{center}
\scalebox{1.0}{
    \begin{tabular}{lrrrr}
    \toprule
    \multicolumn{1}{c}{Project}& \multicolumn{1}{c}{\#total commits} & \multicolumn{1}{c}{\#studied commits} & \multicolumn{1}{c}{\#\oneLine{}s} & \multicolumn{1}{c}{Proportion(\%)}\\
    \midrule
    Camel   & 60,911    & 38,458  & 2,405  &  6.25\\
    Hadoop  & 69,997    & 53,796  & 2,302  &  4.28\\
    Linux   & 1,048,688 & 802,726 & 65,858 &  8.20\\
    Zephyr  & 40,883    & 25,542  & 1,979  &  7.75\\
    \bottomrule
    \end{tabular}
    }
    \end{center}
\end{table}





%% file: codes/motivation/ex-micro-commit.tex

\begin{lstlisting}[caption={An example micro commit in Linux\\retrieved from: 092734b4bb227faddf241b116af14357645d963c}, label={code:ex-micro-commit}, upquote=true, language=diff, basicstyle=\scriptsize, numberstyle=\scriptsize]
@@ -385 +385 @@ EXPORT_SYMBOL(bt878_device_control);
-struct cards card_list[] __devinitdata = {
+static struct cards card_list[] __devinitdata = {
\end{lstlisting}

%% file: codes/motivation/ex-missed-micro-commit.tex

\begin{lstlisting}[caption={An example micro commit with multiple changed lines\\in Linux retrieved from: 0ce6e62bd6591777bd92873e2db93fdbc5228122}, label={code:ex-missed-micro-commit}, upquote=true, language=diff, basicstyle=\tiny, numberstyle=\tiny, escapechar={|}]
@@ -1143,2 +1143,3 @@ int path_lookup_open(const char *name, unsigned int lookup_flags,
-int path_lookup_create(const char *name, unsigned int lookup_flags,
-               struct nameidata *nd, int open_flags, int create_mode)
+static int path_lookup_create(const char *name, unsigned int lookup_flags,
+                             struct nameidata *nd, int open_flags,
+                             int create_mode)
\end{lstlisting}


%% file: sections/dataset.tex
\section{Dataset Preparation}
\label{sec:dataset}

\subsection{Studied Datasets}
To answer our RQs, we conducted an empirical study on four notable large OSS projects written in Java and C: Camel\footnote{\url{https://camel.apache.org/}}, Hadoop\footnote{\url{https://hadoop.apache.org/}}, Linux\footnote{\url{https://www.linux.org/}}, and Zephyr\footnote{\url{https://www.zephyrproject.org/}}.  
The Camel project is an integration framework that provides a routing engine to integrate systems.
The Hadoop project is a distributed computing framework.
The Linux project (a.k.a. the Linux Kernel) is one of the most popular open-source operating system kernels.
The Zephyr project is a real-time operating system supporting several architectures.
Hence, these include four software systems: an integration framework, a distributed computing framework, an operating system kernel, and an operating system. 
We selected these four projects because of three reasons: (1) they are written in popular programming languages (i.e., Java
and C), (2) they are well-known popular OSS projects, and (3) they have a long development history. 


\subsection{One-line Commits and Micro Commits}
\label{sec:database:definition}



Our research aims to accurately quantify small changes using a token-level definition (\ie, micro commits). Additionally, to highlight the differences in accuracy between token-level and line-level definitions, we should compare micro commits with \oneLines{}. Therefore, we detail the process of extracting one-line commits and micro commits from software development histories below.

Git is language agnostic. The changes performed in a commit are displayed as a diff, comparing the code before and after the commit.
These changes are grouped into \emph{hunks}.
A hunk is a set of contiguous lines that are added/removed/modified together, along with metadata that indicates
its context---where the change occurred.
Each hunk can include \emph{context} lines (\ie, lines that were not modified but are used to help interpret the change). 
The default number of context lines is three, but for the purpose of this paper, we have set it to zero; thus, we ignore context lines in the hunk.
Git's diff does not present lines that have been modified. Instead, it simply records lines that
have been removed (prefixed with ``-'') and lines that have been added (prefixed with ``+''); thus, a modified line is represented by a removed
line and its corresponding added line. If several continuous lines are modified simulataneously, Git presents first all
removed lines, and thereafter the added lines.


We extracted \oneLine{}s based on the hunks provided by Git.
Specifically, \oneLine{}s correspond to commits that have a diff with exactly one removed and one added line \textbf{in the same hunk}.
\lis{code:ex-micro-commit} is an example of such a commit.

\input{codes/data-preparation/ex-line-token-commit}

To be able to perform token-level analysis, we processed the repository history using \cregit~\cite{german2019EMSE}.
This uses srcML~\cite{collard2013ICSM} to generate an equivalent commit history where the differences are displayed as changes to sequences of tokens instead of lines (see \cite{german2019EMSE} for a detailed description).
Effectively, we track tokens removed and/or added during a commit and can easily identify commits that have added and/or removed a
certain number of tokens.
Similarly to the way we can identify modified lines, we can identify modified tokens if one token is added and another is removed in the same hunk.
For example, the commit from \lis{code:ex-commit-in-line-repo} is shown in its equivalent token version in
\lis{code:ex-commit-in-token-repo}.

We extracted micro commits based on the hunks provided by Git repositories processed by \cregit.
Micro commits refer to commits that include a maximum of five added tokens and five deleted tokens across all hunks.
This number was chosen for the following reasons.

\begin{itemize}
\item
In the languages being studied (C and Java), it is highly unlikely to add a new statement with only five tokens, suggesting that such commits carry out minor modifications. For example, within five tokens, developers can only add a function call with one parameter and an ending semicolon: \texttt{name(parm);} includes two identifiers, two parentheses, and one semicolon.
\item In the systems we studied, between 7.45 and 17.95\% of all studied commits add at most 5 tokens and remove at most 5 tokens.
\end{itemize}
This number serves as a parameter for micro commits.
For example, we use the same number for both added and deleted tokens while different numbers could be used. 
Its potential threats are discussed in \sec{sec:threats:construct}.




Source code comments are important for source code and making changes to comments are also maintenance activities. However, in this paper, we exclude comments and execute our analysis. The reason is to prioritize maintenance activities for code logic. As mentioned in \sec{sec:introduction}, our intention is to support various software engineering approaches (\eg, defect prediction), which typically prioritize code logic over comments. Indeed, defect prediction studies typically do not take into account comment issues when identifying target defects~\cite{kondo2020EMSE,hoang2019MSR,mcintosh2018TSE}. While we acknowledge the importance of changes made to comments for maintenance purposes, this perspective is beyond the scope of our paper.

\subsection{Data Collection}
\label{sec:dataset:collection}

We preprocessed the commits in the studied repositories and constructed a database with its diffs (both line-based and
token-based) using the following steps.
From this database, we extracted \oneLines{} and micro commits.

\begin{stepitemize}
\item[\textbf{Step 1:}] 
  For each commit, extract the line-based diff of its modified source code ignoring any changes to non-source code:
  \begin{itemize}
  \item Remove changes to non-source files.\footnote{We extract files with the extension of ``java'' in Camel and Hadoop and with the extension of ``c'' and ``h'' in Linux and Zephyr.}
  \item  Remove changes to comments and white space using regular expressions (\eg, ``//.*'').
  \item Remove commits that do not have any changes after the aforementioned processes. 
    \end{itemize}
\item[\textbf{Step 2:}]
   Using \cregit, for each commit, extract the token-based diff of its modified source code ignoring any changes to non-source code:
   \begin{itemize}
   \item Remove changes to comments. 
   \item For each source code token, keep its type and its value. \cregit tokenizes the source code using srcML.\footnote{\url{https://www.srcml.org/}} Thus, the types of tokens are those created by srcML. For example, \texttt{int i;} will be converted
     to the sequence of $type | value$: \verb@name|int, name|i, decl_stml |;@
   \end{itemize}
 \item[\textbf{Step 3:}]
    Create a database in \sqlite with these commits (line and token-based) including:
   \begin{itemize}
   \item Identify and store each hunk and its metadata (such as the file where it occurred and the number of
     lines/tokens added and removed).
   \item Added and removed lines or tokens in each hunk
   \item Commit messages
   \item Metadata (\eg, index)
   \end{itemize}
\end{stepitemize}

In summary, we record for each line-based diff: its commit id and its set of hunks (for each hunk, its location, number of
lines added, number of lines removed, and its contents as a sequence of added/removed lines).
We record the same for token-based diffs (replacing lines with tokens--including their types). 
Note that when obtaining diffs with Git, we use the \emph{myers} algorithm, which is the default algorithm.
Also, we record commit messages.
More details can be found in our replication package (see \sec{sec:threats:external}).

\input{tables/discussion/intersection}

\tab{tab:discussion:intersection} displays the number of extracted micro commits and \oneLine{}s.
We used these micro commits and \oneLine{}s in this study.
We found that micro commits can cover approximately 90\% of \oneLine{}s.
In contract, only approximately 40\% (for Linux and Zephyr) or 50\% (for Camel and Hadoop) of micro commits can be covered by \oneLine{}s. 


\input{tables/rq1/micro-onetoken-commit}
As shown in \tab{tab:micro}, between 7.45\% and 17.95\% of all studied commits are micro commits, and approximately 1 in 3 or 4 micro commits are one-token commits in all projects. 
Hence, micro commits constitute a non-negligible portion of all studied commits.
As expected, most of these commits modify a few lines: between 52.80\% and 58.70\% modify add or remove at most one line, and between 59.56\% and 67.48\% add-or-remove two lines.

%% file: codes/data-preparation/ex-line-token-commit.tex

\begin{lstlisting}[caption={An example commit in a line repository}, label={code:ex-commit-in-line-repo}, upquote=true, language=diff]
@@ -10 +10 @@ test();
-int flg = 10;
+static int flag = 10;
\end{lstlisting}

\begin{lstlisting}[caption={An example commit in a token repository}, label={code:ex-commit-in-token-repo}, upquote=true, language=diff]
@@ -100,2 +100,3 @@ test();
+specifier|static
 name|int
-name|flg
+name|flag
\end{lstlisting}

%% file: tables/discussion/intersection.tex

\begin{table}[t]
  \centering
  \caption{The number and proportion of the intersection between \oneLine{}s and micro commits in each commit type (\ie, one-line or micro).
  The column of ``\#intersects'' indicates the intersection;
  the columns of ``\#one-line'' and ``\#micro'' indicate the number of \oneLine{}s and micro commits;
  the column of ``\%one-line'' and ``\%micro'' indicate the proportion of intersection in each commit type (\ie, \oneLine{}s and micro commits). 
  }
\begin{tabular}{l|r|rr|rr}
\toprule
Project&\#intersects&\#one-line&\#micro&\%one-line&\%micro\\
\midrule
Camel&2,131&2,405&4,230&88.6&50.4\\
Hadoop&2,069&2,302&4,010&89.9&51.6\\
Linux&59,836&65,858&138,142&90.9&43.3\\
Zephyr&1,849&1,979&4,585&93.4&40.3\\
\bottomrule
\end{tabular}
  \label{tab:discussion:intersection}
\end{table}

%% file: tables/rq1/micro-onetoken-commit.tex

\begin{table}[t]
  \centering
  \caption{Number of micro commits and one-token commits and their proportion with respect to all source-code commits.}
\begin{tabular}{l|rr|rr}
\toprule
Project & Micro commits & Prop (\%) & One-token commits & Prop (\%) \\
\midrule
Camel&4,230&11.00&1,319&3.43\\
Hadoop&4,010&7.45&1,288&2.39\\
Linux&138,142&17.21&32,973&4.11\\
Zephyr&4,585&17.95&1,247&4.88\\

\bottomrule


\end{tabular}
  \label{tab:micro}
\end{table}

%% file: sections/rq1.tex
\section{RQ1: \RQCone{}}
\label{sec:rq1}

\subsection{Approach}
The goal of RQ1 is to understand the characteristics of micro commits.
More specifically, we investigated the modified tokens. 

In this RQ, we investigated micro commits from two perspectives: (1) \textit{most frequently modified tokens and token types by micro commits} and (2) \textit{modification patterns for each micro commit}.
We first count added and removed tokens and their token types from all micro commits and provide researchers with tokens and token types frequently modified by micro commits.
Second, we investigate the set of added and removed tokens for every micro commit and show the common \textit{modification patterns} adopted by a single micro commit.
Note that we used the set rather than the sequence of tokens.
Hence, we characterized modification patterns based on modified tokens in micro commits rather than the sequences of modified tokens.

We use srcML classification for the types of tokens. For example,
tokens of type names correspond to names of types and variables (including language predefined ones); literals are constant
values; operators are operators to perform mathematical operations; argument\_list corresponds to either () (empty parameter list), or each of the parenthesis around parameters or the comma that separates them; expr\_stmt is the semicolon at the end of
the statement; block is a \{ or \}; file is a filename; specifier is a C storage specifier (\eg, \texttt{static}); directive a C preprocessor
directive; and annotation corresponds to Java annotations. The right-hand side of C macro definitions is
not further parsed by srcML and is considered a single token of type value (\ie, the value the macro expands to).

\input{figures/rq2/changed-token-type}

\input{figures/rq2/changed-token}

\subsection{Results}

\noindent
\textit{(1) Most frequently modified tokens and token types by micro commits}

\noindent
\textbf{The top-3 most frequently touched token types in micro commits are generally the name, literal, and operator token types.}
\fig{fig:rq2:changed-token-type} shows the frequently added/removed token types by micro commits that account for more than 5\% in all projects.
We found three token types, which we refer to as the top-3 most frequently touched token types.
The token type most frequently included in micro commits is the name token (e.g., the name of a variable or function),
the second one is the literal token (e.g., 123, `a', ``test''), and the third one is the operator
token (such as +).
Also, the proportion of the literal token is significantly different between Java and C.
Specifically, while the proportion of name tokens is more than three times larger than that of literal tokens in the projects written in C, the difference is less than two times in the projects written in Java.
While the proportion of the operator tokens is relatively small, these are also included in the top-3 most frequently touched token types.
Hence, micro commits usually modify name, literal, and operator tokens in most cases, but their proportions may differ between programming languages and their token types.

\textbf{While the tokens corresponding to the top-3 token types differ between Java and C, we observe similar tokens within the same language.}
\fig{fig:rq2:changed-token} shows the top-10 most frequently occurring tokens for the top-3 token types.
In Java, boolean literals (\eg, true/false, null), and numeric literals were the most commonly observed, while in C,
they were the tokens for 0/1, parentheses and names for types (\eg, int, u32\_t, and u8\_t).

In conclusion, the types of tokens most frequently changed are the same in both programming languages, but the actual tokens are different.


\input{tables/rq2/linux-top5-changed-token-types}

\input{codes/rq2/ex-single-token-modification}

\medskip
\noindent
\textit{(2) Modification patterns of micro commits}

\noindent
\textbf{The single token modification is the most frequently observed pattern in the studied micro commits.}
\tab{tab:rq2:frequency-token-types} shows the top-5 most frequently appearing sets of removed and added tokens
in micro commits.
Each row indicates a set of token types modified by a single micro commit and their frequency and proportion (\ie, \# of micro commits). 
The ``n'' column indicates the frequency, while ``Pro'' indicates the proportion. In this paper, we use the same column name in the other tables.
In all projects, the most frequently observed micro commits consist of an added and removed token.
For example, in the Linux project, micro commits adding and removing a name token are the most frequently observed.
This type of single addition and removal usually represents a single token being replaced (\eg,
\lis{code:ex-single-token-modification}) and is the most frequently observed type of micro commit in all projects.


Similar to the results from (1), the modified tokens differ between Java and C. Modifications of literals are the most frequent pattern in Java, accounting for approximately 20\% of all micro commits. Modifications of names are the most common pattern in C, accounting for about 10\% of all micro commits.


\summarybox{Summary of RQ1}
{
	Most micro commits modify a single token, and this token type is either a name, a literal, or an operator.
  The distribution of micro commits of each of these types is different in C and Java.
  The operators being replaced are also significantly different across languages.
}


%% file: figures/rq2/changed-token-type.tex

\begin{figure}[t]
  \centering
  \includegraphics[width=1.00\columnwidth]{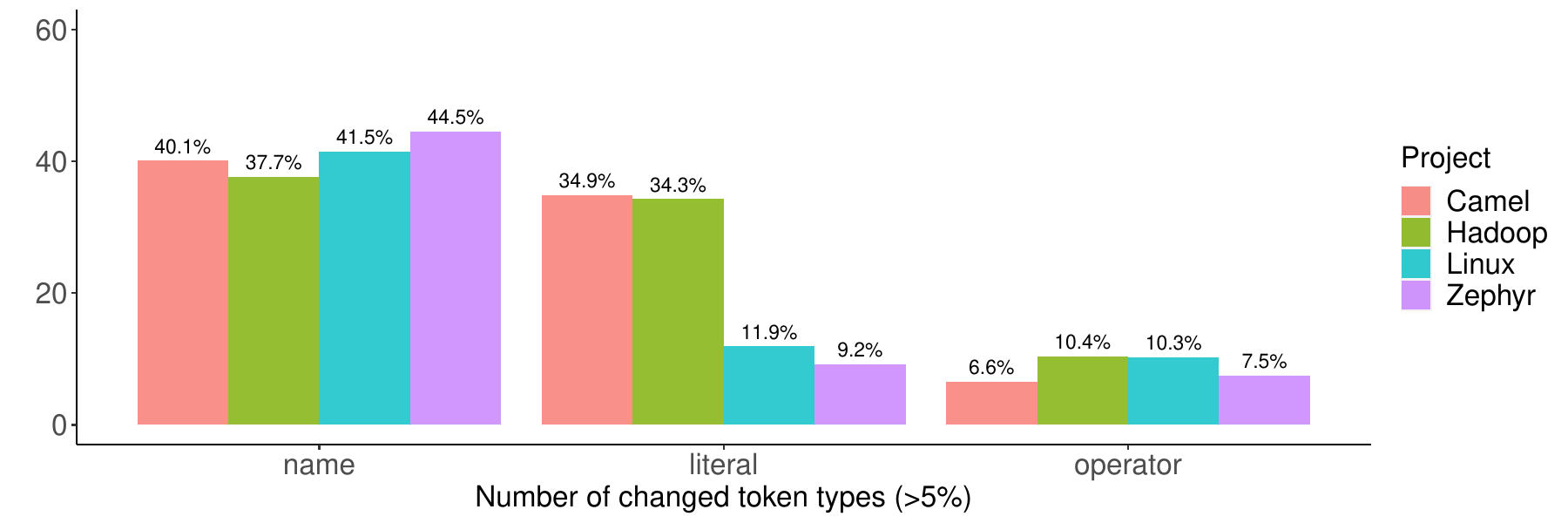}
  \caption{ 
    Proportions of changed token types ($>5\%$)
  }
  \label{fig:rq2:changed-token-type}
\end{figure}

%% file: figures/rq2/changed-token.tex

\begin{figure}[p]
  \centering
  \subfigure[Camel (Java)]{\includegraphics[width=0.7\columnwidth]{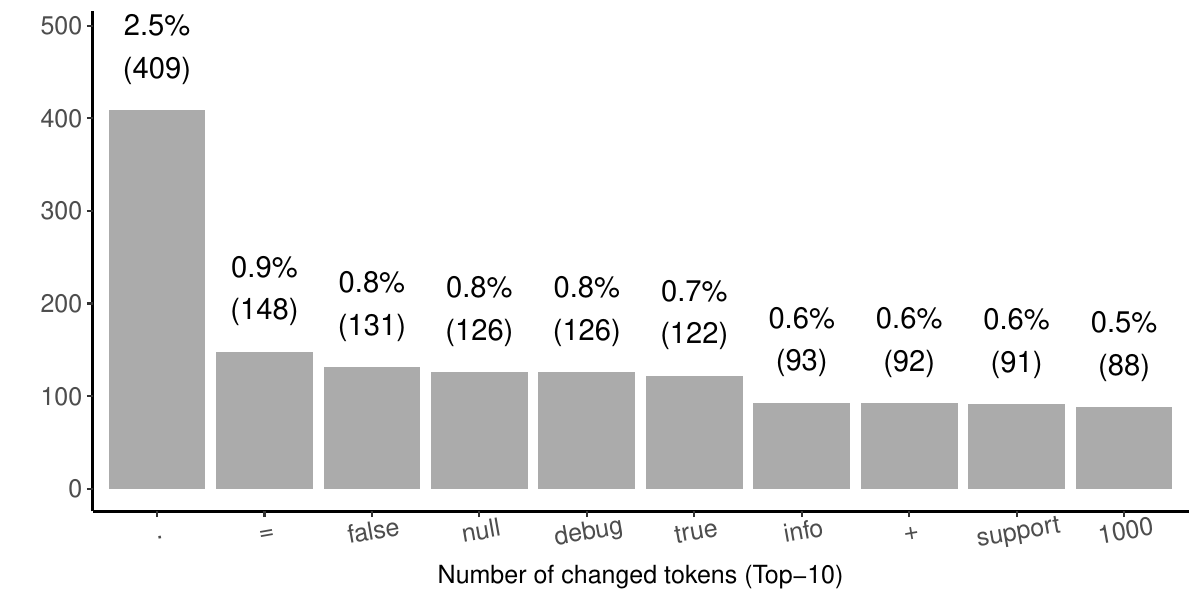}
  \label{fig:rq2:camel-changed-token}}
  \subfigure[Hadoop (Java)]{\includegraphics[width=0.7\columnwidth]{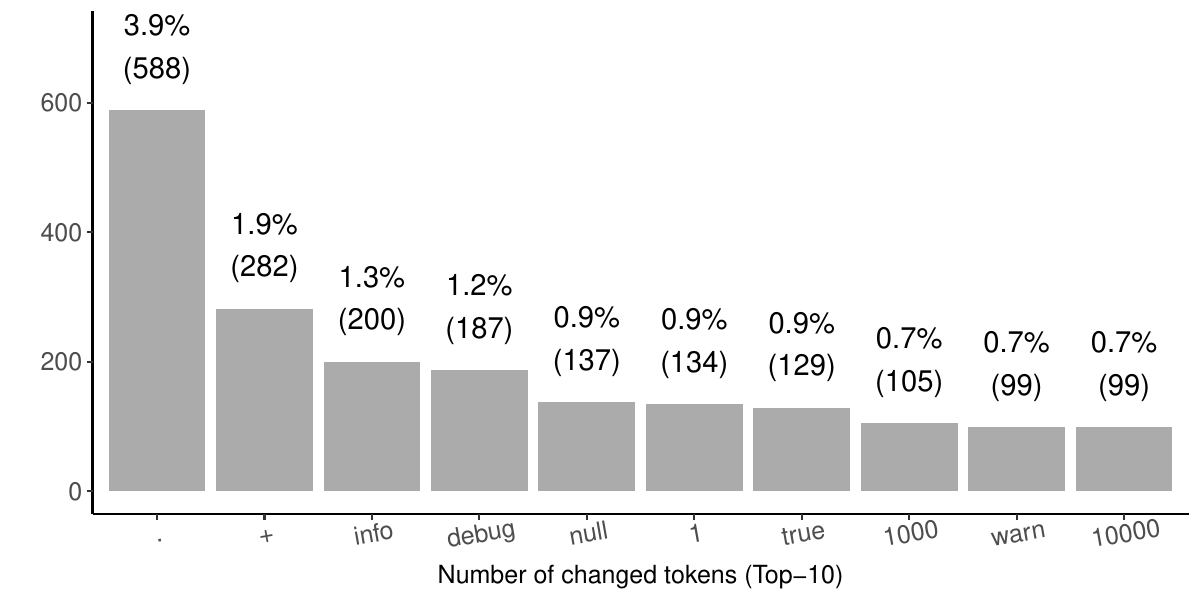}
  \label{fig:rq2:hadoop-changed-token}}
  \subfigure[Linux (C)]{\includegraphics[width=0.7\columnwidth]{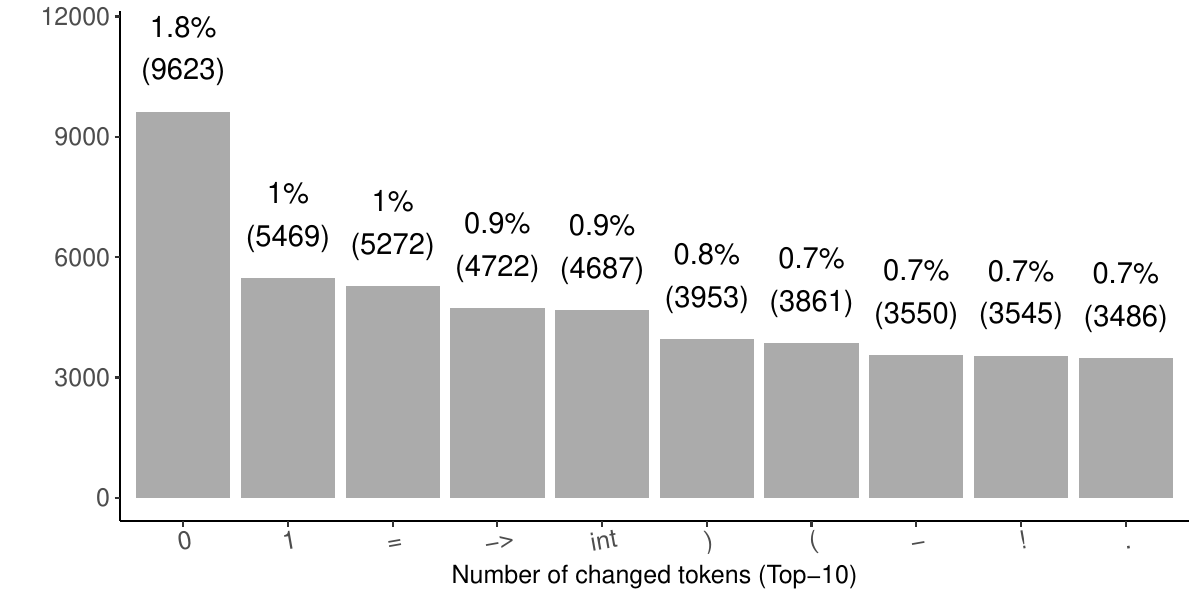}
  \label{fig:rq2:linux-changed-token}}
  \subfigure[Zephyer (C)]{\includegraphics[width=0.7\columnwidth]{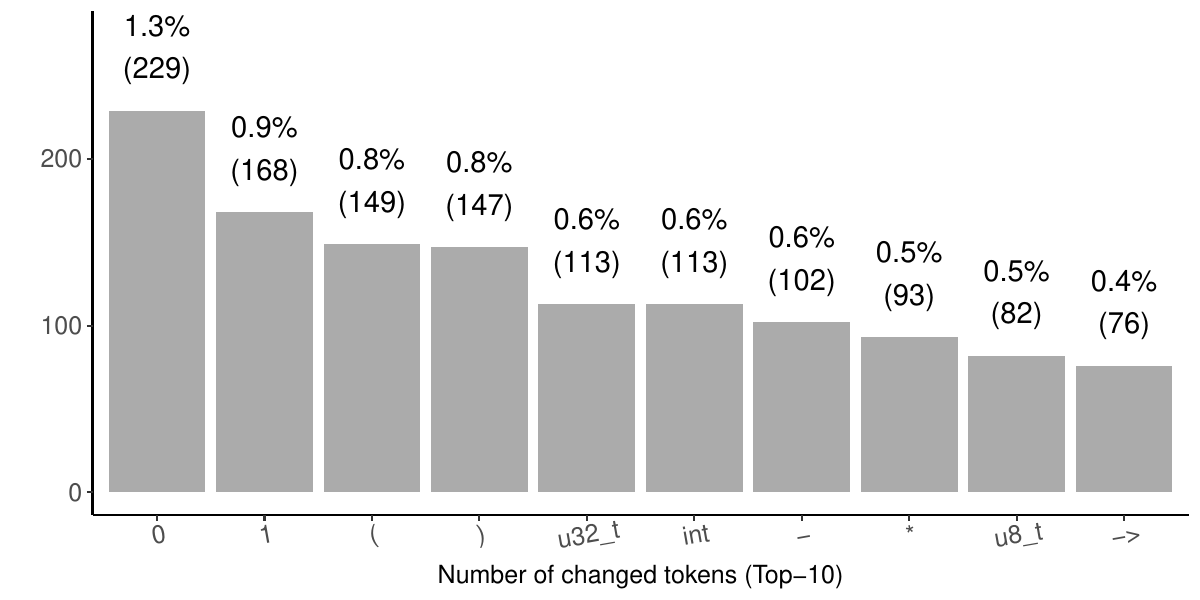}
  \label{fig:rq2:zephyr-changed-token}}
  \caption{ 
    Numbers of changed tokens 
  }
  \label{fig:rq2:changed-token}
\end{figure}

%% file: tables/rq2/linux-top5-changed-token-types.tex

\begin{table}[t]
    \caption{
        Top-5 added and removed token types applied to source code by micro commits
}
\label{tab:rq2:frequency-token-types}
\begin{center}
\scalebox{0.9}{
    \begin{tabular}{lllrr}
    \toprule
    \multicolumn{1}{c}{Project} & \multicolumn{1}{c}{Add} & \multicolumn{1}{c}{Remove} & \multicolumn{1}{c}{n} & \multicolumn{1}{c}{Pro}                                                                                \\
    \midrule
                        &literal         &literal         &833   &\Chart{19.7}{0.197}\\
\multirow{2}{*}{Camel}  &name            &name            &435   &\Chart{10.3}{0.103}\\
\multirow{2}{*}{(Java)} &literal,literal &literal,literal &312   &\Chart{7.4}{0.074}\\
                        &name,name       &name,name       &269   &\Chart{6.4}{0.064}\\
                        &name,name,name  &name,name,name  &125   &\Chart{3.0}{0.030}\\
    \midrule
                        &literal         &literal         &761   &\Chart{19.0}{0.190}\\
\multirow{2}{*}{Hadoop} &name            &name            &408   &\Chart{10.2}{0.102}\\
\multirow{2}{*}{(Java)} &literal,literal &literal,literal &266   &\Chart{6.6}{0.066}\\
                        &name,name       &name,name       &229   &\Chart{5.7}{0.057}\\
                        &specifier       & -              &126   &\Chart{3.1}{0.031}\\
    \midrule
                        &name            &name            &13693 &\Chart{9.9}{0.099}\\
\multirow{2}{*}{Linux}  &literal         &literal         &7350  &\Chart{5.3}{0.053}\\
\multirow{2}{*}{(C)}    &value           &value           &6835  &\Chart{4.9}{0.049}\\
                        &name,name       &name,name       &6141  &\Chart{4.4}{0.044}\\
                        &specifier       & -              &4171  &\Chart{3.0}{0.030}\\
    \midrule
                        &name            &name            &487   &\Chart{10.6}{0.106}\\
\multirow{2}{*}{Zephyr} &value           &value           &352   &\Chart{7.7}{0.077}\\
\multirow{2}{*}{(C)}    &name,name       &name,name       &260   &\Chart{5.7}{0.057}\\
                        &name,name,name  &name,name,name  &170   &\Chart{3.7}{0.037}\\
                        &literal         &literal         &169   &\Chart{3.7}{0.037}\\

    \bottomrule
    \end{tabular}
    }
    \end{center}
\end{table}

%% file: codes/rq2/ex-single-token-modification.tex
\begin{lstlisting}[caption={An example single token modification in a token repository}, label={code:ex-single-token-modification}, upquote=true, language=diff]
@@ -100,1 +100,1 @@ test();
-name|flg
+name|flag
\end{lstlisting}

%% file: sections/rq2.tex
\section{RQ2: \RQCtwo{}}
\label{sec:rq2}

\subsection{Approach}

This RQ aims to understand the details of the activities performed by micro commits.
Specifically, we manually inspect a large set of micro commits to understand what types of change were performed from a source code perspective, considering removed and/or added tokens.
Such an understanding gives us insight into whether understanding micro commits can support several approaches in software engineering (see the details in \sec{sec:discussion}).

Our manual inspection consisted of two phases: (1) \textit{constructing a coding guide} and (2) \textit{manual classification}. 
Constructing a coding guide for manual classification/annotation is a common practice in the field of mining software repositories~\cite{hata2019ICSE,wang2021EMSE,chouchen2021SANER,hata2022EMSE,wang2023EMSE,zanaty2018ESEM}. To create the coding guide, we referred to previous studies~\cite{hata2019ICSE,wang2021EMSE,chouchen2021SANER,hata2022EMSE,wang2023EMSE,zanaty2018ESEM} and followed the process detailed below.

The initial coding guide was first discussed by the first and second authors. Since this is the first study to classify micro commits, we examined both micro commits and other types of commits to develop the initial coding guide. After constructing the initial coding guide, we aimed to reach a consensus among the first three authors for this guide and refine the guide. Specifically, we independently annotated 20 micro commits from a subset of all micro commits. This subset consists of micro commits that only change less than or equal to five tokens in main files (.c or .java files) in the Linux, Hadoop, and Zephyr projects to investigate a single operation commit for refining the coding guide.
We computed the agreement rate for these 20 micro commits using Fleiss' Kappa~\cite{fleiss1971} that is used to demonstrate inter-rater agreement when there are more than two raters. It is also frequently applied in the field of mining software repositories~\cite{chouchen2021SANER,hata2022EMSE}.
The Kappa coefficient is commonly interpreted using the following scale~\cite{viera2005}:
Slight agreement ($0.01 \leq k \leq 0.20$),
Fair agreement ($0.21 \leq k \leq 0.40$),
Moderate agreement ($0.41 \leq k \leq 0.60$),
Substantial agreement ($0.61 \leq k \leq 0.80$),
Almost perfect agreement ($0.81 \leq k \leq 0.99$).
Then we discussed the coding guide along with any inconsistencies in categorization to reach a consensus. We repeated this process until our categorization substantially matched, indicating that our coding guide was successfully constructed. We, therefore, repeated this process three times (\ie, independently classifying 60 commits).
Finally, our agreement rate achieved substantial agreement in two consecutive iterations.
\tab{tab:rq3:final-kappa} shows all agreement rates across three authors for each repetition.

\input{tables/rq3/kappa.tex}

Through this process, we identified two perspectives: \emph{operation} and \emph{target}.
The operation indicates what kind of operations are applied, such as adding a new statement or changing an expression; the target indicates source code entities where the operation is applied, such as expressions.
The details of the coding guide are described below.

We utilized the following coding guide to categorize micro commits in terms of the operations.
\begin{itemize}
  \item \emph{add:} This refers to operations that add a completely new entity.
  \item \emph{replace:} This refers to operations that modify an entity.
  \item \emph{remove:} This refers to operations that completely remove an entity.
  \item \emph{multi:} This code indicates that multiple operations are applied.
  \item \emph{no:} This code indicates that no functional change is applied.
\end{itemize}

We utilized the following coding guide to categorize micro commits in terms of the targets.
\begin{itemize}
  \item \emph{identifier:} This refers to commits that only modify identifiers, such as variable names. If other entities, such as parentheses, are included, it would not be labelled as an identifier but would be considered an expression.
  \item \emph{statement:} This refers to commits that modify a complete statement, including the semicolon (;), such as an entire function call with its semicolon. C's \#include preprocessor statement is also regarded as a statement.
  \item \emph{constant:} This refers to commits that only modify literals, such as strings or numbers. If other entities, such as parentheses, are included, they would not be identified as a constant but would be considered an expression.
  \item \emph{declaration:} This refers to commits that modify declarations, such as variable declarations. However, if the commit can be classified as ``identifier'' or ``constant'', it should be categorized under these two categories rather than ``declaration''.
  \item \emph{control flow:} This refers to commits that modify the control flow of execution, such as adding a new ``else'' statement.
  \item \emph{expression:} If a commit does not match other categories and involves modifying a part of a statement, it would be classified into this category. Additionally, this category includes transformations from constants to variables or vice versa, as well as conversions from a variable to a pointer and vice versa.
  \item \emph{multi:} This code suggests that operations are performed on multiple targets.
  \item \emph{no:} This code indicates that no functional change is applied.
\end{itemize}

\input{tables/rq3/manual-coding-result.tex}
\input{codes/rq3/linux-ex1.tex}
\input{codes/rq3/linux-ex2.tex}

\tab{tab:rq3:manual-coding-result} shows the summary and examples.
This represents different types of activities performed by micro commits.
Let us describe two example commits. Listing~\ref{code:rq3-ex1} shows an example commit. This commit changes a function call and its argument.
More specifically, the identifiers of the function call and the argument value are replaced so that we classify this commit as operations$=$replace, and targets$=$identifier.
The commit of Listing~\ref{code:rq3-ex2} replaces an expression ``$++$'' into ``$--$''.  Hence, we classify this commit as operations$=$replace, and targets$=$expression.

Because our agreement rates for operation and target based on our coding guide achieved almost perfect and substantial agreement respectively (\tab{tab:rq3:final-kappa}) and we made an internal consensus of the coding guide, only the first author manually classified the 400 micro commits, similar to previous studies~\cite{wang2021EMSE,wang2023EMSE,hata2019ICSE,hata2022EMSE}. 
All manual categorizations are available in our sheet in our replication package.\footnote{\url{https://doi.org/10.5281/zenodo.10963270}}
The sample size in manual inspection was determined as a statistical representative with a confidence level of 95\% and a confidence interval of 5\% for 150,967 micro commits from all studied projects.\footnote{\url{https://www.surveysystem.com/sscalc.htm}}
The minimum sample size with this confidence level and this confidence interval is 383. For safety, we also inspect 17 additional micro commits. Therefore, we classify a total of 400 micro commits.

\subsection{Results}

\input{tables/rq3/single-multi-matrix}

\input{tables/rq3/single-activity}

\input{codes/rq3/linux-ex4}

\textbf{Micro commits usually perform a single operation (85.75\%). Micro commits with multiple operations account for the remaining 14.25\%, and they usually correspond to two operations.}
\tab{tab:rq3:sin-mul-matrix} lists the proportion of micro commits classified into ``multi'' in the operation and target.
Approximately 86\% of micro commits are classified into single operations (\ie, non ``multi''); thus, micro commits usually modify an extremely small section.
We refer to such commits as \textit{single-operation micro commits} (\eg, Listing~\ref{code:ex-single-token-modification} is a replace identifier).
We surprisingly observe that approximately 14\% are classified into \textit{multi-operation micro commits} (\ie, ``multi'').
Hence, multi-operation micro commits account for a non-negligible portion of the micro commits.
Listing~\ref{code:rq3-ex4} shows an example of a multi-operation micro commit.
This commit has an add declaration and a replace constant. 
We disregard the first hunk because it only contains comment lines.
It should be noted that such multi-operation micro commits usually have two operations only; we observed that only two commits contain more than two operations.
Thus, based on our manual inspection, it is rare to find micro commits containing more than two operations.


\textbf{Approximately 82.2\% of single-operation micro commits replace existing tokens.}
\tab{tab:rq3:single-activity} summarizes the frequency of the combination of operations and targets in single-operation micro commits.
The top-4 combinations include the ``replace'' operation, accounting for approximately 78.7\% (270/343).
Also, all the ``replace'' operation commits account for 82.2\% (282/343). 
This result suggests that many single-operation micro commits modify the existing source code, but do not add or remove the source code.

\textbf{Multi-operation micro commits more frequently add-and-remove statements rather than single-operation ones.}
\tab{tab:rq3:multi-unique-activity} summarizes the frequency of the combination of operations and targets in multi-operation micro commits, and \tab{tab:rq3:multi-activity} summarizes the frequency of the pair of their combination for each commit.
The main difference from single-operation micro commits is that the ``add statement'' and the ``remove statement'' are top-3 (\tab{tab:rq3:multi-unique-activity}).
The reason is that multi-operation micro commits that add and remove statements appear most frequently (\tab{tab:rq3:multi-activity}).
This type of commit is used to move the statements, and therefore, 
change the order of execution and potentially the control flow of the program.
For example, Listing~\ref{code:rq3-ex3} shows an example micro commit.
This commit adds and removes a statement and changes the order of execution of the free statement to fix a bug related to freeing data.
This swapping activity is frequently observed in multi-operation micro commits. 

\input{tables/rq3/multi-activity-unique}

\input{tables/rq3/multi-activity}

\input{codes/rq3/linux-ex3}

\summarybox{Summary of RQ2}
{
More than 85\% of micro commits apply a single operation to a single target, and they mainly replace existing target.
Multi-operation micro commits frequently change the order of statements. 
}


%% file: tables/rq3/kappa.tex

\begin{table}[t]
    \caption{
     Fleiss' Kappa scores for each repetition
}
\label{tab:rq3:final-kappa}
\begin{center}
\scalebox{1.0}{
    \begin{tabular}{lccc}
    \toprule
		\multicolumn{1}{c}{Criteria} & \multicolumn{1}{c}{First Time}&\multicolumn{1}{c}{Second Time}&\multicolumn{1}{c}{Third Time}\\
    \midrule
    Operations & 0.686 & 0.669 & 0.832 \\
    Targets    & 0.425 & 0.671 & 0.754 \\
    \bottomrule
    \end{tabular}
    }
    \end{center}
\end{table}

%% file: tables/rq3/manual-coding-result.tex

\begin{table*}[t]
    \caption{
      The description of each candidate in our manual inspection
}
\label{tab:rq3:manual-coding-result}
\begin{center}
    \begin{tabular}{llp{2cm}p{6cm}}
    \toprule
		\multicolumn{1}{c}{Criteria} & \multicolumn{1}{c}{Candidate} & \multicolumn{1}{c}{Description} & \multicolumn{1}{c}{Example Commits and Their Diffs in Linux}               \\
    \midrule
                                             & add                           & Add a new entity                                    & \texttt{122503683169b21d9cdb90380a20caad7ba994b6} Diff: \lis{code:app-operation-add}\\
\cmidrule(l){2-4}                              
                                             & replace                       & Replace an existing entity                                     & \texttt{b7a90e8043e7ab1922126e1c1c5c004b470f9e2a} Diff: \lis{code:app-operation-replace}\\
\cmidrule(l){2-4}                              
    Operations                               & remove                        & Remove a completely existing entity                            & \texttt{b95b4e1ed92a203f4bdfc55f53d6e9c2773e3b6d} Diff: \lis{code:app-operation-remove}\\
\cmidrule(l){2-4}                              
                                             & multi                         & Operations on multiple targets                               & \texttt{8df0cfe6c6c4a9355989baa8de9f166b2bc51f76} Diff: \lis{code:rq3-ex4}\\
\cmidrule(l){2-4}                              
                                             & no                            & Non-functional modification                           & \texttt{a092532483e3200a53c8b1170b3988cc668c07ef} Diff: \lis{code:app-operation-no}\\
    \midrule
                                             & declaration                   & Change in a type signature                 & \texttt{36f062042b0fd9f8e41b97a472f52139886ca26f} Diff: \lis{code:app-entity-declaration}\\
\cmidrule(l){2-4}                              
                                             & constant                      & A constant (\eg, literal)                                      & \texttt{1db76c14d215c8b26024dd532de3dcaf66ea30f7} Diff: \lis{code:app-entity-constant}\\
\cmidrule(l){2-4}                              

                                             & identifier                    & An identifier (\eg, function calls)                            & \texttt{70e8b40176c75d3544024e7c934720b11a8a11bf} Diff: \lis{code:app-entity-identifier}\\
\cmidrule(l){2-4}                              

     \multirow{2}{*}{Targets}                & control flow                  & Modifies the control flow                                           & \texttt{415a1975923722f729211a9efca550c60c519bf3} Diff: \lis{code:app-entity-controlflow}\\
\cmidrule(l){2-4}                              
                                             & statement                     & A the majority of a statement (delimited
                                                                               by semicolon)                              & \texttt{b95b4e1ed92a203f4bdfc55f53d6e9c2773e3b6d} Diff: \lis{code:app-operation-remove}\\
\cmidrule(l){2-4}                              
                                             & expression                    & A part of a statement and not classified into other categories & \texttt{40cc394be1aa18848b8757e03bd8ed23281f572e} Diff: \lis{code:app-entity-expression}\\
\cmidrule(l){2-4}                              
                                             & multi                         & Multiple targets are altered                                  & \texttt{8df0cfe6c6c4a9355989baa8de9f166b2bc51f76} Diff: \lis{code:rq3-ex4}\\
\cmidrule(l){2-4}                              
                                             & no                            & Non-functional modification                           & \texttt{a092532483e3200a53c8b1170b3988cc668c07ef} Diff: \lis{code:app-operation-no}\\
    \bottomrule
    \end{tabular}
    \end{center}
\end{table*}


%% file: codes/rq3/linux-ex1.tex

\begin{lstlisting}[caption={Example ``replace identifier'' commit diff retrieved from\\
  \protect\texttt{f72e6c3e17be568138d8e4855ac2734d251a6913} in Linux.}, label={code:rq3-ex1}, upquote=true, language=diff, basicstyle=\scriptsize, numberstyle=\scriptsize]
-       strlcpy(drvinfo->bus_info, pci_name(mdev->pdev),
+       strlcpy(drvinfo->bus_info, dev_name(mdev->device),
\end{lstlisting}

%% file: codes/rq3/linux-ex2.tex

\begin{lstlisting}[caption={Example ``replace expression'' commit diff retrieved from\\
  \protect\texttt{8b58f261113c442717b9d205ab187e51c3823597} in Linux.}, label={code:rq3-ex2}, upquote=true, language=diff]
-       dqm->total_queue_count++;
+       dqm->total_queue_count--;
\end{lstlisting}

%% file: tables/rq3/single-multi-matrix.tex

\begin{table}[t]
    \caption{
     Proportion of micro commits having multi activities
}
\label{tab:rq3:sin-mul-matrix}
\begin{center}
\scalebox{1.0}{
    \begin{tabular}{cc}
    \toprule
		\multicolumn{1}{c}{Single}&\multicolumn{1}{c}{Multi}\\
    \midrule
    85.75\%(343)&14.25\%(57)\\
    \bottomrule
    \end{tabular}
    }
    \end{center}
\end{table}

%% file: tables/rq3/single-activity.tex

\begin{table}[t]
    \caption{
      The frequency of the combination of operations and targets in single-operation micro commits 
}
\label{tab:rq3:single-activity}
\begin{center}
\scalebox{1.0}{
    \begin{tabular}{llrr}
    \toprule
    \multicolumn{1}{c}{Operation}&\multicolumn{1}{c}{Target}&\multicolumn{1}{c}{n}&\multicolumn{1}{c}{Pro}\\
    \midrule
    replace&expression&85&\Chart{24.8}{0.248}\\
    replace&identifier&69&\Chart{20.1}{0.201}\\
    replace&constant&59&\Chart{17.2}{0.172}\\
    replace&declaration&57&\Chart{16.6}{0.166}\\
    add&statement&22&\Chart{6.4}{0.064}\\
    replace&control flow&12&\Chart{3.5}{0.035}\\
    no&no&8&\Chart{2.3}{0.023}\\
    remove&statement&8&\Chart{2.3}{0.023}\\
    add&expression&7&\Chart{2.0}{0.020}\\
    remove&declaration&7&\Chart{2.0}{0.020}\\
    add&control flow&4&\Chart{1.2}{0.012}\\
    add&identifier&2&\Chart{0.6}{0.006}\\
    remove&control flow&1&\Chart{0.3}{0.003}\\
    remove&expression&1&\Chart{0.3}{0.003}\\
    remove&identifier&1&\Chart{0.3}{0.003}\\
    \bottomrule
    \end{tabular}
    }
    \end{center}
\end{table}

%% file: codes/rq3/linux-ex4.tex

\begin{lstlisting}[caption={Example ``multi'' micro commit diffs retrieved from\\
  \protect\texttt{8df0cfe6c6c4a9355989baa8de9f166b2bc51f76} in Linux.}, label={code:rq3-ex4}, upquote=true, language=diff, basicstyle=\scriptsize, numberstyle=\scriptsize]
@@ -111,0 +112,5 @@
+ * - EXTCON_PROP_USB_SS (SuperSpeed)
+ * @type:       integer (intval)
+ * @value:      0 (USB/USB2) or 1 (USB3)
+ * @default:    0 (USB/USB2)
+ *
@@ -114,0 +120 @@
+#define EXTCON_PROP_USB_SS             2
@@ -117 +123 @@
-#define EXTCON_PROP_USB_MAX            1
+#define EXTCON_PROP_USB_MAX            2
\end{lstlisting}

%% file: tables/rq3/multi-activity-unique.tex

\begin{table}[t]
    \caption{
      The frequency of the combination of operations and targets in multi-operation micro commits 
}
\label{tab:rq3:multi-unique-activity}
\begin{center}
\scalebox{1.0}{
    \begin{tabular}{llrr}
    \toprule
    \multicolumn{1}{c}{Operation}&\multicolumn{1}{c}{Target}&\multicolumn{1}{c}{n}&\multicolumn{1}{c}{Pro}\\
    \midrule

    replace&identifier&22&\Chart{19.1}{0.191}\\
    replace&expression&19&\Chart{16.5}{0.165}\\
    add&statement&16&\Chart{13.9}{0.139}\\
    remove&statement&16&\Chart{13.9}{0.139}\\
    replace&constant&15&\Chart{13.0}{0.130}\\
    add&expression&7&\Chart{6.1}{0.061}\\
    replace&declaration&5&\Chart{4.3}{0.043}\\
    remove&expression&4&\Chart{3.5}{0.035}\\
    add&declaration&3&\Chart{2.6}{0.026}\\
    replace&control flow&3&\Chart{2.6}{0.026}\\
    add&control flow&2&\Chart{1.7}{0.017}\\
    remove&declaration&2&\Chart{1.7}{0.017}\\
    remove&control flow&1&\Chart{0.9}{0.009}\\

    \bottomrule
    \end{tabular}
    }
    \end{center}
\end{table}

%% file: tables/rq3/multi-activity.tex

\begin{table}[t]
    \caption{
      The frequency of the pair of the combination of operations and targets in multi-operation micro commits 
}
\label{tab:rq3:multi-activity}
\begin{center}
\scalebox{1.0}{
    \begin{tabular}{llllllrr}
    \toprule
    \multicolumn{1}{c}{Ope\#1}&\multicolumn{1}{c}{Tar\#1}&\multicolumn{1}{c}{Ope\#2}&\multicolumn{1}{c}{Tar\#2}&\multicolumn{1}{c}{Ope\#3}&\multicolumn{1}{c}{Tar\#3}&\multicolumn{1}{c}{n}&\multicolumn{1}{c}{Pro}\\
    \midrule
    remove&statement&add&statement&-&-&11&\Chart{19.3}{0.193}\\
    replace&identifier&replace&expression&-&-&6&\Chart{10.5}{0.105}\\
    replace&identifier&replace&constant&-&-&6&\Chart{10.5}{0.105}\\
    add&expression&replace&identifier&-&-&4&\Chart{7.0}{0.070}\\
    replace&expression&replace&constant&-&-&4&\Chart{7.0}{0.070}\\
    remove&statement&replace&expression&-&-&3&\Chart{5.3}{0.053}\\
    add&expression&replace&constant&-&-&2&\Chart{3.5}{0.035}\\
    add&control flow&replace&expression&-&-&2&\Chart{3.5}{0.035}\\
    replace&identifier&replace&declaration&-&-&2&\Chart{3.5}{0.035}\\
    replace&expression&replace&control flow&-&-&2&\Chart{3.5}{0.035}\\
    add&declaration&remove&declaration&-&-&2&\Chart{3.5}{0.035}\\
    add&statement&remove&expression&replace&identifier&2&\Chart{3.5}{0.035}\\
    replace&constant&remove&expression&-&-&1&\Chart{1.8}{0.018}\\
    add&statement&replace&identifier&-&-&1&\Chart{1.8}{0.018}\\
    remove&statement&replace&control flow&-&-&1&\Chart{1.8}{0.018}\\
    add&statement&replace&declaration&-&-&1&\Chart{1.8}{0.018}\\
    replace&declaration&replace&constant&-&-&1&\Chart{1.8}{0.018}\\
    add&statement&replace&expression&-&-&1&\Chart{1.8}{0.018}\\
    add&statement&remove&control flow&-&-&1&\Chart{1.8}{0.018}\\
    add&declaration&replace&constant&-&-&1&\Chart{1.8}{0.018}\\
    remove&statement&replace&identifier&-&-&1&\Chart{1.8}{0.018}\\
    add&expression&remove&expression&-&-&1&\Chart{1.8}{0.018}\\
    replace&declaration&replace&expression&-&-&1&\Chart{1.8}{0.018}\\

    \bottomrule
    \end{tabular}
    }
    \end{center}
\end{table}

%% file: codes/rq3/linux-ex3.tex

\begin{lstlisting}[caption={Example ``multi'' commit diff retrieved from\\
  \protect\texttt{a71bfb4a6aabfe5e6f145883020153103c7fdfba} in Linux.}, label={code:rq3-ex3}, upquote=true, language=diff]
-error_free_data:
-       free(data);
 error_free_buffer_access:
        free(buffer_access);
+error_free_data:
+       free(data);
\end{lstlisting}

%% file: sections/rq3.tex
\section{RQ3: \RQCthree{}}
\label{sec:rq3}

\subsection{Approach}


As discussed in Sections~\ref{sec:introduction} and~\ref{sec:motivation}, \oneLines{} are common in software development and often address deficiencies in the system~\cite{purushothaman2005TSE}. However, they have a drawback: they overlook changes within a line. Consequently, two commits with the same number of changed tokens could differ; one might be a \oneLine{} and the other might not.
The concept of micro commits, introduced in this paper, address their drawbacks.


However, the extent of the differences between one-line commits and micro commits is unclear.
Extracting micro commits is more costly than one-line commits as it requires syntactic parsing of the source code.
Hence, this RQ aims to compare the one-line commits and micro commits.

Our methodology can be summarized as follows: we start by analyzing the changed tokens in one-line commits.
We then analyze the modified hunks in micro commits.
Finally, we discuss the intersection between \oneLines{} and micro commits. 

\subsection{Results}

\input{figures/rq1/heatOneLineTokens}
\input{figures/rq1/token1lAccum}

\textbf{Approximately 90\% of one-line commits consist of at most five tokens.}
\tab{tab:motivation:micro-commits} shows the number of \oneLines.
As described in \sec{sec:motivation}, there are a non-negligible number of these commits in the studied projects (4.28--8.20\%).
Figure~\ref{fig:oneLineHeat} shows the proportion of \oneLines{} according to the number of tokens that they
have added and removed between 0 and 10.
As can be seen, there are a significant number of \oneLines that remove and add exactly one token (between approximately 50 and 63\% of all \oneLine{}s).
Furthermore, except for the case in the Hadoop project where no commits add or delete five tokens, all cells with five or fewer added and deleted tokens have more than one \oneLine{} across all projects.  
This implies that there are no empty cells within five added or deleted tokens except for one cell in the Hadoop project.
Also, the distribution of \oneLines{}, with more than five tokens, varies across the projects. For instance, in the Hadoop and Zephyr projects, there are cells with no \oneLines{} of more than five deleted tokens and less than or equal to one added token.
In contrast, every cell in the Camel and Linux projects has at least one \oneLine{}.
Hence, the majority of \oneLine{}s add or remove at most five tokens, and this finding is generally consistent across all projects.

Figure~\ref{fig:oneLineAccum} shows the accumulated distribution of \oneLines according to the maximum number of
tokens they add or remove.
We use the maximum number of tokens added or removed in this figure.
This is because our definition of a micro commit applies the same threshold of five tokens to both the number of added and removed tokens.
As can be seen, between approximately 57\% and 65\% add-or-remove at most one token, between 76\% and
82\% add-or-remove at most three tokens, and between 89\% and 93\% add-or-remove at most five tokens.
Thus, approximately 90\% of \oneLines{} can be covered by our micro commits.

\input{figures/rq1/num_hunks_linerepo}

The number of modified hunks is also a crucial characteristic of commits. By our definition, \oneLine{}s only modify one location in the source code (\ie, one hunk). We define micro commits based on the number of tokens, so even if a commit is spread across multiple locations (\ie, multiple hunks), it it still considered a micro commit if the number of modified tokens is below a certain threshold. This is a significant distinction compared to \oneLine{}s. Therefore, we do not impose any limits on the number of modified hunks.

\fig{fig:num_hunk_linerepo} illustrates the accumulated distribution of the number of hunks included in micro commits to investigate their difference from \oneLine{}s.
Approximately 70\% (Linux and Hadoop) or 60\% (Zephyr and Camel) of micro commits contain a single hunk, while the remaining commits encompass two or more hunks. Hence, while approximately 70\% or 60\% of micro commits share characteristics with \oneLine{}s, the remaining 30\% or 40\% represent commits that one-line commits do not detect, even if they modify the same number of tokens.

In conclusion, although micro commits can encompass nearly all \oneLines, the reverse is not typically true: \oneLines do not generally cover micro commits.
Indeed, \tab{tab:discussion:intersection} in \sec{sec:dataset:collection} reveals that around 90\% of one-line commits can be encapsulated by micro commits. However, only approximately 40\% (for Linux and Zephyr) or 50\% (for Camel and Hadoop) of micro commits can be encapsulated by one-line commits.
Therefore, micro commits provide new insights compared to one-line commits.

\summarybox{Summary of RQ3}
{
  Approximately 90\% of \oneLines add or remove at most five tokens. 
  Therefore, nearly all one-line commits can be covered by micro commits.
  In contrast, 30 to 40\% of micro commits include two or more hunks that are not covered by one-line commits. In fact, only approximately 40\% (for Linux and Zephyr) or 50\% (for Camel and Hadoop) of micro commits can be encapsulated by one-line commits. 
  Therefore, the characteristics of micro commits can help us understand the attributes of small changes, including those in one-line commits and commits not identified by one-line commits.
}


%% file: figures/rq1/heatOneLineTokens.tex

\begin{figure*}[htbp]
  \centering
      \includegraphics[width=0.95\textwidth]{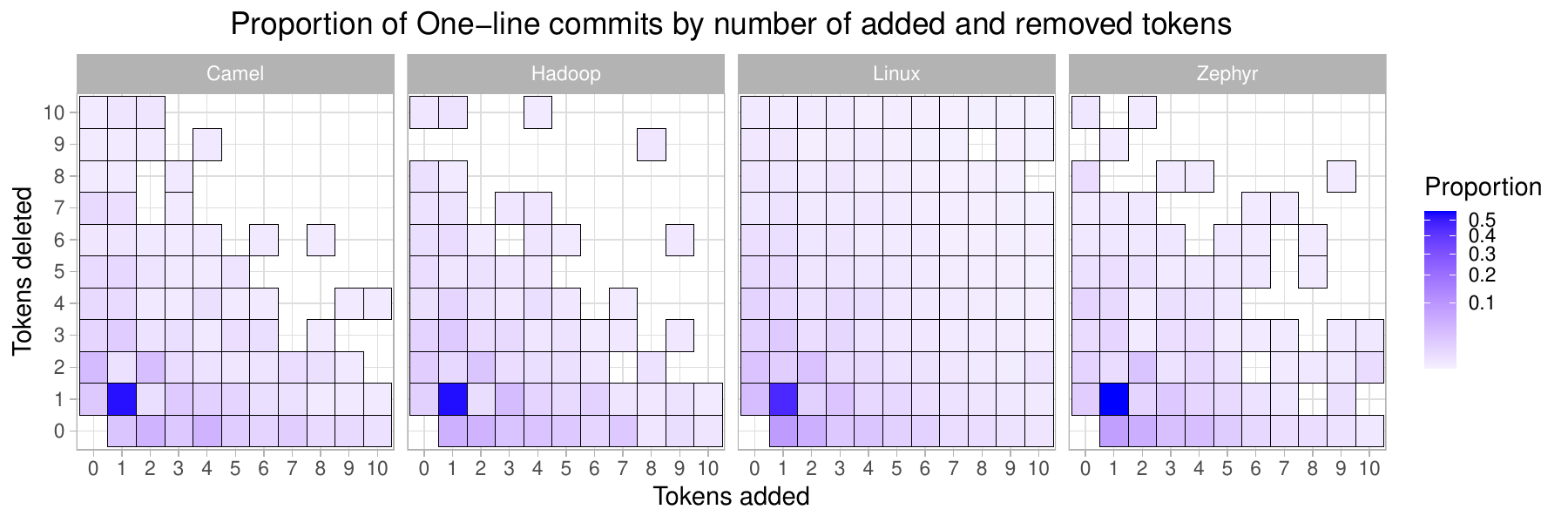}
  \caption{Proportion of \textbf{\oneLines} by the number of tokens added or removed. The x and y-axis show the added and deleted tokens, and each cell indicates the proportion of commits.}
  \label{fig:oneLineHeat}
\end{figure*}

%% file: figures/rq1/token1lAccum.tex

\begin{figure}[htbp]
  \centering
      \includegraphics[width=0.8\textwidth]{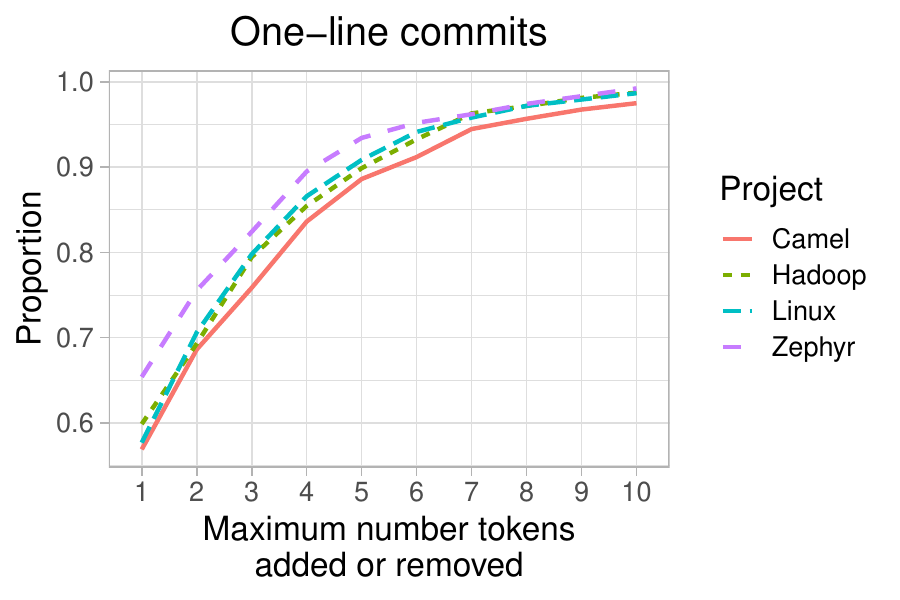}
  \caption{Accumulated distribution of \oneLine{}s in terms of the maximum number of added or removed tokens}
  \label{fig:oneLineAccum}
\end{figure}

%% file: figures/rq1/num_hunks_linerepo.tex

\begin{figure}[htbp]
  \centering
      \includegraphics[width=0.8\textwidth]{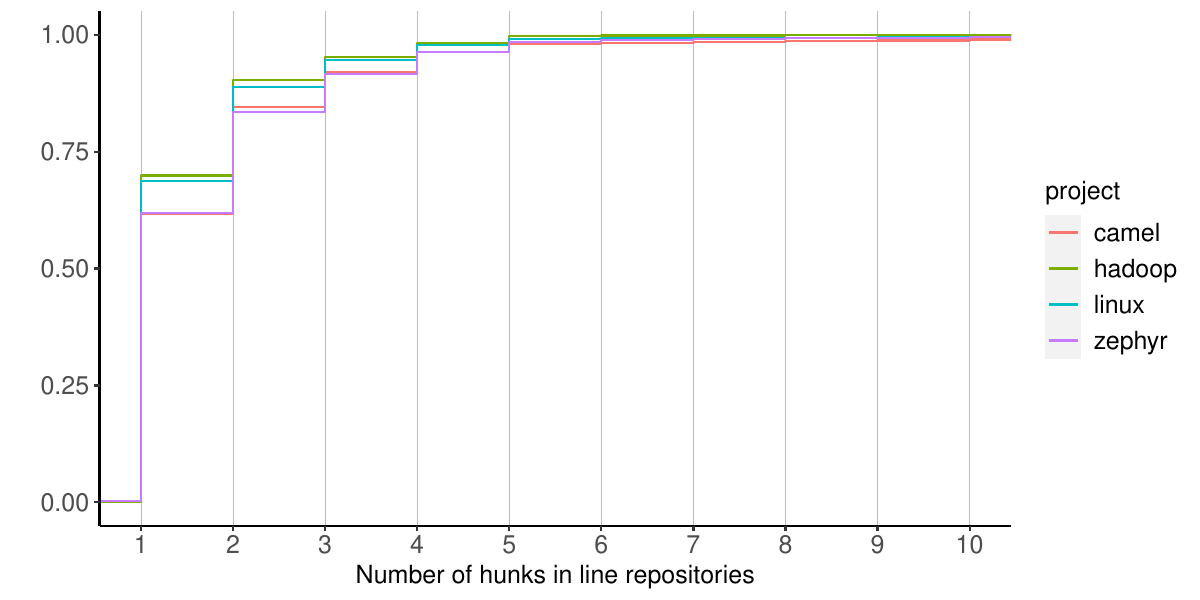}
  \caption{Accumulated distribution of micro commits ($N=5$) in terms of the number of hunks included}
  \label{fig:num_hunk_linerepo}
\end{figure}

%% file: sections/discussion.tex
\section{Discussion}
\label{sec:discussion}

In this paper, we defined micro commits and shed light on their characteristics.
The main motivation is that turning our attention to micro commits would benefit software engineering research.
In this section, we describe the implications of micro commits on future research.

\subsection{Line-based vs Token-based Complexity Metrics}
Line-based complexity metrics (\eg, LA, LD, and LT)~\cite{kamei2013TSE} are one of the most popular source code
complexity metrics in software engineering. 
However, \textbf{there are
  limitations to using line-based complexity metrics, as they may overlook capturing commits with equivalent complexity
  in terms of changed tokens.}  We have shown examples with different numbers of changed lines but the same number of
changed tokens (\liss{code:ex-micro-commit} and~\ref{code:ex-missed-micro-commit}).  We have shown that in the projects
under analysis, approximately 90\% of one-line commits are micro commits, but only approximately 40--50\% of micro commits are one-line commits (\tab{tab:discussion:intersection}).

Thus, \textbf{future research should consider tokens (and their types) as an additional
metric of the complexity of commits}. 
Because, for example, prior studies~\cite{kondo2020EMSE,mcintosh2018TSE} in defect prediction reported that current models heavily rely on the added lines, such new metrics would provide new information to identify defective commits accurately.

\subsection{Micro Commits Are Non-negligible and Should Be Further Studied}

In \sec{sec:dataset:collection}, we showed that micro commits account for between 7.45 and 17.95\% of all studied commits, which is
quite high. Furthermore, 1 in 3 or 4 these changes (2.39 and 4.88\% of all studied commits) simply change one token.
Hence, micro commits, including their finest-grained form, the one-token commit, represent a non-negligible development activity.
Thus, \textbf{we need to understand how to better support developers, first, by deeply looking at the need for micro commits, and second, by reducing the amount of effort needed to complete these changes}.

\subsection{Program Repair}
\label{sec:discussion:apr}

The results of RQ1 showed that micro commits frequently modify a single token, and its token type is name, literal, or operator.
\textbf{Studying micro commits could help understand how software is modified with such a tiny amount of change, and provide datasets that improve methods that attempt to modify software automatically.}
For example, datasets based on micro commits might improve data-driven program repair approaches that have been studied so far~\cite{jiang2018ISSTA,liu2018ICSME,martinez2015EMSE}. 
One potential idea involves utilizing our observations of frequently modified token types and tokens in Java and C.
Our observations indicate that while the types of frequently modified tokens are similar, the actual tokens differ across languages.
This information is important for developing a program repair approach.
When dealing with multiple languages, focusing on token types is crucial.
However, when focusing on a specific language, actual tokens can also be beneficial.

\input{codes/discussion/change-string-literal}
Also, \textbf{the empirical investigation of micro commits would reveal types of micro commits that are difficult to be generated by program repair approaches}. 
Listing~\ref{code:change-string-literal} shows a \emph{replace constant} micro commit example.
Specifically, this commit changes a string literal token: ``start'' into ``end''.
Such a change might be difficult to be generated automatically, because it is not obvious why a literal token should be replaced
by another one; however, other changes (including micro commits) might have performed this specific replacement
somewhere else.

Some large commits might actually be composed of several micro commits (\ie, \textit{tangled commits}~\cite{dias2015SANER,herzig2013MSR,kirinuki2014ICPC}). Thus \textbf{it is worth also exploring the
possibility of untangling micro commits from larger commits}. These untangled micro commits might be very valuable for
program repair.

Finally, we present initial analysis results for micro commits regarding their maintenance activities. As stated in \sec{sec:introduction}, we hypothesize that small changes are likely intended for maintenance purposes. Therefore, we deduce that exploring micro commits could be beneficial for program repair. To validate this hypothesis, we identify micro commits that fall under the corrective maintenance category as defined by Swanson~\cite{swanson1976ICSE}. Corrective maintenance is performed in response to failures. If corrective maintenance makes up a large proportion of micro commits compared to non-micro commits, it would confirm our hypothesis.

To identify the corrective commits, we followed the methodologies used in prior studies~\cite{mockus2000ICSM,purushothaman2005TSE}, which use keywords in commit messages.
More specifically, if at least one of the keywords is included in the commit messages, we classify the commit into the corrective maintenance category.
Otherwise, we do not label commits. 
We used the keyword list defined by Levin~\et~\cite{levin2016ICSME} as follows:
``fix'', ``esolv'', ``clos'', ``handl'', ``issue'', ``defect'', ``bug'', ``problem'', ``ticket''.

The detailed procedure is as follows.
\begin{stepitemize}
\item[\textbf{Step 1:}] 
  Apply preprocessing to the commit messages using the NLTK package\footnote{\url{https://www.nltk.org/}} in Python by following the steps below:
  \begin{itemize}
    \item
    Tokenize the text and convert all words to lowercase.
    \item
    Remove stopwords and punctuation.
    \item
    Perform stemming on all words.
  \end{itemize}
\item[\textbf{Step 2:}]
   Check if the stemmed commit message contains a keyword.
 \item[\textbf{Step 3:}]
    Identify commits that fall under the corrective category.
\end{stepitemize}

\input{figures/discussion/swanson-categorization}

\textbf{Micro commits are more likely to be failure-fixing activity than other commits.}
\fig{fig:rq3:swanson} shows the proportion of corrective micro and non-micro commits.
In this figure, we compare the tendency of micro commits (light gray) and non-micro commits (dark gray).
Corrective micro commits are larger than non-micro commits.
Hence, micro commits distinguishably correspond to the corrective commits.
This result shows that micro commits are usually applied to the source code to fix failures. 

Also, this finding confirms our initial assumption that micro commits are used more frequently for maintenance purposes than non-micro commits.
Interestingly, Hattori and Lanza~\cite{hattori2008ASE} found similar results, noting that tiny commits are often associated with corrective activities.

It should be noted that the keyword-based approach generally lacks accuracy~\cite{alomar2021JSS,kondo2022ILA}.
To verify the accuracy of the identification, we manually inspect 20 micro commits and 20 non-micro commits identified as corrective, classifying them into three failure types within the corrective category defined by Swanson~\cite{swanson1976ICSE}. If we cannot associate any failure types, those would be considered false positive corrective commits. This allows us to estimate the actual number of corrective micro commits and corrective non-micro commits in the identified commits. We do not examine non-corrective commits to determine the proportion of false-negative corrective commits. This discussion only reports the minimum percentage of corrective commits. Our manual inspection revealed that there were no false-positive corrective micro commits.
In contrast, we found 8 out of 20 false positive corrective commits in non-micro commits.
This finding suggests that the percentage of corrective micro commits may not change significantly, while the proportion of corrective non-micro commits could decrease. Therefore, our conclusion remains unchanged.
In this manual inspection, we inspect only 20 micro and non-micro commits. Future studies could improve the validity of our findings.
Our inspection is avaiable in our sheet in our replication package.\footnote{\url{https://doi.org/10.5281/zenodo.10963270}}


\subsection{Size-perspective vs. Semantic-perspective for Defining Micro Commits}
In this paper, we define micro commits through size metrics  (\ie, the number of tokens). This is because we would like to assist with software engineering research, such as program repair. However, micro is a general term, and micro commits can be defined not only by size but also by semantic aspects. For instance, tangled commits~\cite{dias2015SANER,herzig2013MSR,kirinuki2014ICPC} can be considered non-micro, whereas non-tangled commits can be categorized as micro. Additionally, defect-fixing commits can be categorized as micro or non-micro depending on the difficulty of the bug being fixed. We could explore these aspects using non-source code resources, such as source code comments, issue reports, and mailing lists. Exploring these semantic-based micro commits can also contribute to software engineering research.


%% file: codes/discussion/change-string-literal.tex

\begin{lstlisting}[caption={Example ``replace constant'' commit diff retrieved from\\
  \protect\texttt{c143708acfb17e91c5e4fc9bd9b496fc7d2db29c} in Hadoop.}, label={code:change-string-literal}, upquote=true, language=diff, basicstyle=\tiny, numberstyle=\tiny]
@@ -71 +71 @@ protected void render(Block html) {
-      html.h1()._("Invalid log start value: " + $("end"))._();
+      html.h1()._("Invalid log end value: " + $("end"))._();
\end{lstlisting}

%% file: figures/discussion/swanson-categorization.tex

\begin{figure*}[t]
  \centering
  \includegraphics[width=1.0\columnwidth]{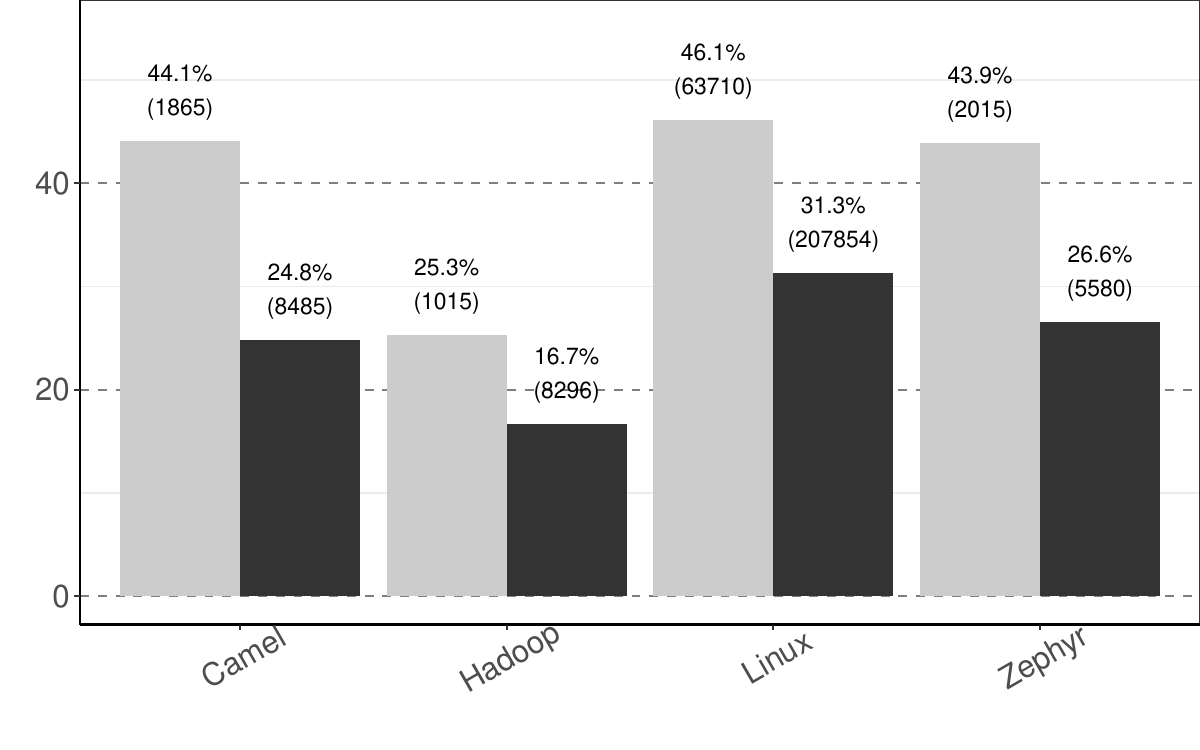}
  \caption{ 
    The proportion of commits for the corrective category. Light gray indicates the proportion in micro commits; dark gray indicates the proportion in non-micro commits. 
  }
  \label{fig:rq3:swanson}
\end{figure*}

%% file: sections/threats.tex
\section{Threats to Validity}
\label{sec:threats}

\subsection{External Validity}
\label{sec:threats:external}
We conducted our empirical study on four OSS projects.
To mitigate the threats to generalizability, we selected OSS projects that are active, popular, and well-known OSS projects written in two popular programming languages. However, even if we use these OSS projects, our results may not be generalized to all projects.
Indeed, these are system software.
To remedy this challenge, replication studies in research or practical scenarios (e.g., actual projects in the industry) are necessary. Hence, we provide a replication package.\footnote{\url{https://doi.org/10.5281/zenodo.10963270}}
Also, the key tool \cregit is an OSS tool; thus, researchers and practitioners easily convert their Git repositories into token-based ones.

\subsection{Construct Validity}
\label{sec:threats:construct}
We define micro commits based on the number of changed tokens. However, micro commits are a general term, and we can make different definitions. The key characteristic of micro commits is that such commits change a small code fragment. Our analysis (RQ1 and 2) shows that our definition is consistent with this characteristic.
Hence, we believe our definition can be acceptable. However, our definition may not be the best; thus, future studies are necessary to find a better definition than our first one. 
For example, future studies can investigate different thresholds for the number of tokens. Also, studying different thresholds for added and deleted tokens (\eg, 3 added tokens and 5 deleted tokens) can be beneficial. Additionally, they can consider changes to source code comments. This definition would encompass not only maintenance activities related to code logic, but also various other maintenance activities.

For future studies, researchers can use ASTs to tokenize the source code instead of \cregit{}, which we used in this paper.
While ASTs are powerful in analyzing token-level information, \cregit{} is designed for Git, a de facto standard version control system.
Hence, researchers easily analyze the software development history to support its process when using \cregit{} instead of ASTs.
Hence, we recommend using \cregit{} in future studies.   

There are several factors that can influence commits. For instance, the way developers write commits can vary depending on the developer and the project. To reduce these influences, we chose and analyzed four projects that involve a large number of developers. As a result, we expect the impact of such influences has been minimized.

\subsection{Internal Validity}
To remove comment lines from source code files, we use regular expressions.
This process is not perfect and may overlook comment lines. 
However, our manual analysis in RQ2 observes that our regular expressions usually work well because we do not find false-positive micro commits.
Therefore, we can reduce the risks associated with using regular expressions.

Also, another threat exists in our manual analysis (RQ2). In this analysis, we performed manual labeling to micro commits according to our coding guide. Because this process is performed manually by the first author, the result may have false-positive and false-negative results.
Therefore, we have made all labels publicly available to facilitate the validation of future studies.
Also, to construct the coding guide, the first three authors independently inspected 20 micro commits three times. This process may also include errors. However, our agreement rate achieved substantial agreement in two consecutive iterations. Hence, we believe the coding guide is reliable.
An alternative solution is to use an automatic classification approach rather than manual analysis. We developed a heuristic-based method to classify micro commits into their corresponding targets automatically.
Overall, this approach achieved an accuracy of approximately 81.2\% for categorizing micro commits into their targets. However, classifying micro commits corresponding to multiple targets can be challenging, with an accuracy of approximately 10.5\%.
To facilitate replication of this approach, we have included it in our replication package.
Finally, we randomly sampled 400 micro commits from all projects. Therefore, our sampled micro commits may be biased by the size of the original projects. To mitigate this risk, we manually inspected additional micro commits from each project. We do not observe significant differences across the projects.

In the discussion, we use keywords to identify the commits related to the corrective maintenance activity as defined by Swanson~\cite{swanson1976ICSE}.
While the keyword identification is widely used to categorize commits~\cite{purushothaman2005TSE,mockus2000ICSM,levin2016ICSME,levin2017PROMISE,hindle2008MSR,karampatsis2020MSR}, it is not perfect~\cite{alomar2021JSS,kondo2022ILA}.
To mitigate this threat, we manually review identified commits and estimate their accuracy.
Also, there are other sets of maintenance activities that can be used to classify commits, such as the IEEE standard~\cite{IEEEStandard}.
While we believe the maintenance activities defined by Swanson are acceptable, future studies are necessary to use other sets. 
Also, if commit messages do not contain any keywords, we exclude those commits from the analysis. However, it is possible that these commits are related to maintenance activities. 
Using more precise methods would enhance the validity of this analysis.

The tool ``cregit'' used to tokenize the source code files utilizes srcML. Therefore, our analysis can only be applied to specific versions of Java (Java SE8 Edition) and C (up to C11) that are supported by srcML. We can find the supported versions on the official homepage.\footnote{\url{https://www.srcml.org/#home}}
To extend our analysis to different versions of Java and C, it is necessary to update srcML and apply our analysis to those versions.

%% file: sections/related.tex
\section{Related Work}
\label{sec:related}

\subsection{Challenges of Mining Git Repositories}
Prior studies~\cite{german2019EMSE,spacco2009IEEE,asaduzzaman2013ICSM,bird2009MSR} investigated and intended to address the challenges of mining Git repositories. 
For example, as described in~\sec{sec:motivation}, some non-functional changes update the information for each line and make it difficult to track code changes accurately.
\cregit~\cite{german2019EMSE} is proposed to address this problem by improving the blame feature in Git.
More specifically, \cregit tokenizes each line and applies the blame feature to the tokenized files. 
Spacco and Williams~\cite{spacco2009IEEE} proposed a technique SDiff to track changes at the statement level instead of the line level. 
This technique combines previous line- and structural-based approaches. 
Specifically, SDiff tokenizes each statement and uses diff between revisions. 
These techniques tokenize the source code to address this problem.
Similarly, we define micro commits based on changed tokens in this paper to track code changes accurately.

\subsection{Change Classification}
Classifying changes (\eg, commits) into a certain category is a research topic in mining software repositories so far~\cite{purushothaman2005TSE,mockus2000ICSM,levin2016ICSME,levin2017PROMISE,hindle2008MSR,karampatsis2020MSR,ghadhab2021IST,hindle2009ICPC,alali2008ICPC,williams2010IST,german2006EMSE,mauczka2015MSR,yan2016JSS}. 
For example, many prior studies intend to classify changes in terms of the purpose~\cite{mockus2000ICSM,levin2016ICSME,levin2017PROMISE,hindle2008MSR,karampatsis2020MSR,ghadhab2021IST,hindle2009ICPC,german2006EMSE,mauczka2015MSR,yan2016JSS}.
Levin~\et~\cite{levin2016ICSME} classified commits into the maintenance activities defined by Swanson~\cite{swanson1976ICSE}. 
Hindle~\et~\cite{hindle2009ICPC} used machine learning classifiers to classify changes into the extended Swanson categories. 
Ghadhab~\et~\cite{ghadhab2021IST} used a pre-trained deep learning model known as BERT to classify commits into maintenance categories. 

On the contrary, in this paper, we classify commits into micro commits based on their size and empirically investigate their characteristics, and there are several similar prior studies~\cite{purushothaman2005TSE,hindle2008MSR,alali2008ICPC}.
Purushothaman and Perry~\cite{purushothaman2005TSE} classified changes into three categories and studied them: one-line changes, small changes, and all. Specifically, this study used the number of changed lines for this classification. 
Hindle~\et~\cite{hindle2008MSR} identified large commits based on the number of changed files and revealed the characteristics of large commits.
They also compared their result with the characteristics of the small commits by Purushothaman and Perry~\cite{purushothaman2005TSE}.  
Alali~\et~\cite{alali2008ICPC} empirically investigated the characteristics of commits in nine OSS projects.
They used three size criteria: the number of files, lines, and hunks.
For example, they found that approximately 19.9\% of commits in the GNU gcc system change at most five lines. 
However, the finest-grained changed source code entity is a line in these papers, and such an entity loses the information of changed tokens in a line.
This limitation makes it difficult to define a certain category of commits based on finer-grained source code changes.
Hence, our investigation would provide a new research direction in which researchers and practitioners use token-level changes.


\subsection{Knowledge Gap in Previous Studies}
Compared to these prior studies, this research is the first to define micro commits at a fine granularity, specifically at the token level, through empirical analysis.
Small commits defined at the line level, which previous studies often used, may overlook important information for improving existing software engineering research. Our research addresses this knowledge gap by conducting the analysis at the token level.

For instance, as explained in \sec{sec:discussion:apr}, the findings of RQ1 in this study have implications for research in program repair. These findings indicate the need to explore approaches for fixing bugs caused by a single name or literal token. This is because existing automated program repair approaches~\cite{goues2012TSE,liu2019ISSTA,jiang2018ISSTA} may not be effective in such scenarios due to a lack of information to repair the code.
These findings and implications were obtained because the analysis was conducted at the token level. It would have been difficult to obtain such findings and implications using a line-level analysis.  The novelty of this study lies in conducting the analysis at the token level and providing these implications.
The details of our findings and implications can be found in Sections~\ref{sec:rq2}, \ref{sec:rq3} and \ref{sec:discussion}.



%% file: sections/conclusion.tex
\section{Conclusion}
\label{sec:conclusion}

In this paper, we defined micro commits (add at most five tokens and remove at most five tokens) and investigated their characteristics.
This research is the first to define micro commits at a fine granularity, specifically at the token level.
The key novelty of this study lies in conducting the analysis at the token level and providing implications for software engineering research.

Below, we present a summary of the findings from our empirical analysis:
\begin{itemize}
	\item 
	Our defined micro commits account for between 7.45--17.95\% of all studied commits.
	Approximately 1 in 3 or 4 these changes (2.39--4.88\% of all studied commits) involve replacing one token with another.  
	Furthermore, RQ3 demonstrates that approximately 90\% of \oneLine{}s are micro commits, but only approximately 40--50\% of micro commits are \oneLine{}s. In fact, approximately 30--40\% of micro commits include two or more hunks. 
	\item
	The results of RQ1 show that micro commits primarily affect name token types (37.7--44.5\%), literal token types (9.2--34.9\%), or operator token types (6.6--10.4\%). The most frequently affected tokens vary: the period in Java (2.5\% in Camel and 3.9\% in Hadoop) and the 0/1 in C (1.8 and 1.0\% in Linux and 1.3 and 0.9\% in Zephyr).
	Furthermore, the most frequently observed pattern is the modification of a single token. In Java projects, this modification is typically a single literal token. On the other hand, in C projects, the modification is usually a single name token.
	\item
	The results of RQ2 indicate that approximately 86\% of micro commits involve a single operation on a single target, with the main focus being the replacement of existing targets. The multi-operation micro commits primarily involve changing the order of statements (19.3\%).
\end{itemize}

In the discussion, we presented the following four implications of micro commits on future research based on the findings:
\begin{itemize}
	\item 
	Based on RQ3, it is observed that almost all \oneLine{}s are micro commits, whereas only 40--50\% of the micro commits are \oneLine{}s. Therefore, token-based complexity metrics offer supplementary information to the commonly used line-based complexity metrics. Designing metrics to measure token-based complexity is a potential area for future research.
	\item
	Based on the statistics of micro commits, they account for a non-negligible proportion of all studied commits (7.45--17.95\%). Additionally, according to \sec{sec:discussion:apr}, these commits are more likely used to fix bugs. Therefore, supporting the development of micro commits is an important area for future research.
	\item
	Based on RQ1, micro commits frequently modify a single token, with the token type often being either a name or a literal. While these micro commits often address bug fixes, suggesting patches to fix individual name or literal tokens can be challenging with existing program repair approaches. Therefore, it is necessary to investigate these micro commits and propose new program repair approaches for future research.
	\item
	We define micro commits based on size metrics. However, micro is a general term, and micro commits can be defined not only by size but also by semantic aspects (\eg, tangled commits or not). Exploring semantic-based micro commits is a potential area for future research.
\end{itemize}

The key message of this paper is as follows:

\begin{quote}
	\emph{The token-level definition could help researchers and practitioners to improve software engineering approaches for software quality assurance activities.}
\end{quote}


%% file: sections/appendix.tex
%
%

\appendix
\begin{Large}
	\noindent \textbf{Appendix}
\end{Large}

\normalsize
\vspace{-1ex}

\section{Example micro commits}

\input{codes/appendix/operation-add}
\input{codes/appendix/operation-replace}
\input{codes/appendix/operation-remove}
\input{codes/appendix/operation-no}
\input{codes/appendix/entity-declaration}
\input{codes/appendix/entity-constant}
\input{codes/appendix/entity-identifier}
\input{codes/appendix/entity-controlflow}
\input{codes/appendix/entity-expression}

%% file: codes/appendix/operation-add.tex

\begin{lstlisting}[float, floatplacement=h, caption={Example ``add statement'' commit diff retrieved from\\
  \protect\texttt{122503683169b21d9cdb90380a20caad7ba994b6} in Linux.}, label={code:app-operation-add}, upquote=true, language=diff, basicstyle=\scriptsize, numberstyle=\scriptsize]
@@ -240,0 +241 @@ static int check_partial_mapping(struct drm_i915_gem_object *obj,
+               cond_resched();
\end{lstlisting}

%% file: codes/appendix/operation-replace.tex

\begin{lstlisting}[float, floatplacement=h, caption={Example ``replace constant'' commit diff retrieved from\\
  \protect\texttt{b7a90e8043e7ab1922126e1c1c5c004b470f9e2a} in Linux.}, label={code:app-operation-replace}, upquote=true, language=diff, basicstyle=\scriptsize, numberstyle=\scriptsize]
@@ -174 +174 @@ static int hfsplus_sync_fs(struct super_block *sb, int wait)
-       dprint(DBG_SUPER, "hfsplus_write_super\n");
+       dprint(DBG_SUPER, "hfsplus_sync_fs\n");
\end{lstlisting}

%% file: codes/appendix/operation-remove.tex

\begin{lstlisting}[float, floatplacement=h, caption={Example ``remove statement'' commit diff retrieved from\\
  \protect\texttt{b95b4e1ed92a203f4bdfc55f53d6e9c2773e3b6d} in Linux.}, label={code:app-operation-remove}, upquote=true, language=diff, basicstyle=\tiny, numberstyle=\tiny]
@@ -2973 +2972,0 @@ void * __meminit alloc_pages_exact_nid(int nid, size_t size, gfp_t gfp_mask)
-EXPORT_SYMBOL(alloc_pages_exact_nid);
\end{lstlisting}

%% file: codes/appendix/operation-no.tex

\begin{lstlisting}[float, floatplacement=h, caption={Example ``no'' commit diff retrieved from\\
  \protect\texttt{a092532483e3200a53c8b1170b3988cc668c07ef} in Linux.}, label={code:app-operation-no}, upquote=true, language=diff, basicstyle=\tiny, numberstyle=\tiny]
@@ -1035 +1035 @@ static int __dwc3_gadget_ep_queue(struct dwc3_ep *dep, struct dwc3_request *req)
-       };
+       }
\end{lstlisting}

%% file: codes/appendix/entity-declaration.tex

\begin{lstlisting}[float, floatplacement=h, caption={Example ``change declaration'' commit diff retrieved from\\
  \protect\texttt{36f062042b0fd9f8e41b97a472f52139886ca26f} in Linux.}, label={code:app-entity-declaration}, upquote=true, language=diff, basicstyle=\scriptsize, numberstyle=\scriptsize]
@@ -382 +382 @@ static ssize_t read_vmcore(struct file *file, char __user *buffer,
-static int mmap_vmcore_fault(struct vm_fault *vmf)
+static vm_fault_t mmap_vmcore_fault(struct vm_fault *vmf)
\end{lstlisting}

%% file: codes/appendix/entity-constant.tex

\begin{lstlisting}[float, floatplacement=h, caption={Example ``change constant'' commit diff retrieved from\\
  \protect\texttt{1db76c14d215c8b26024dd532de3dcaf66ea30f7} in Linux.}, label={code:app-entity-constant}, upquote=true, language=diff, basicstyle=\scriptsize, numberstyle=\scriptsize]
@@ -185 +185 @@ struct mthca_cmd_context {
-static int fw_cmd_doorbell = 1;
+static int fw_cmd_doorbell = 0;
\end{lstlisting}

%% file: codes/appendix/entity-identifier.tex

\begin{lstlisting}[float, floatplacement=h, caption={Example ``change identifier'' commit diff retrieved from\\
  \protect\texttt{70e8b40176c75d3544024e7c934720b11a8a11bf} in Linux.}, label={code:app-entity-identifier}, upquote=true, language=diff, basicstyle=\scriptsize, numberstyle=\scriptsize]
@@ -564 +564 @@ static irqreturn_t pcie_isr(int irq, void *dev_id)
-                       return IRQ_HANDLED;
+                       return IRQ_NONE;
\end{lstlisting}

%% file: codes/appendix/entity-controlflow.tex

\begin{lstlisting}[float, floatplacement=h, caption={Example ``change control flow'' commit diff retrieved from\\
  \protect\texttt{415a1975923722f729211a9efca550c60c519bf3} in Linux.}, label={code:app-entity-controlflow}, upquote=true, language=diff, basicstyle=\scriptsize, numberstyle=\scriptsize]
@@ -530,0 +531 @@ static int ir_probe(struct i2c_adapter *adap)
+               break;
\end{lstlisting}

%% file: codes/appendix/entity-expression.tex

\begin{lstlisting}[float, floatplacement=h, caption={Example ``change expression'' commit diff retrieved from\\
  \protect\texttt{40cc394be1aa18848b8757e03bd8ed23281f572e} in Linux.}, label={code:app-entity-expression}, upquote=true, language=diff, basicstyle=\scriptsize, numberstyle=\scriptsize]
@@ -1922 +1922 @@ static int nfs_parse_devname(const char *dev_name,
-                       *comma = 0;
+                       len = comma - dev_name;
\end{lstlisting}